\documentclass[journal,10pt]{IEEEtran}
\usepackage{dblfloatfix}
\ifCLASSINFOpdf
\else
\fi
\usepackage{amsfonts,epsfig, cite, array, multirow,graphicx,amsmath,ltablex,tabularx,setspace,amssymb, multirow}
\usepackage{schemabloc,tikz,amsthm}
\usepackage{algorithm}
\usepackage{algorithmic}
\usepackage{array}
\usepackage[caption=false,font=normalsize,labelfont=sf,textfont=sf]{subfig}
\usepackage[small]{caption}
\usepackage[numbers,sort&compress]{natbib}
\usetikzlibrary{circuits, arrows}
\usepackage{verbatim}
\usepackage{graphicx}
\usepackage{epsfig}
\usepackage{balance}
\usepackage{epstopdf}
\usepackage{graphicx}
\usepackage{amsmath}
\usepackage{enumitem}
\usepackage{amssymb}
\usepackage{graphics}
\usepackage{cite}
\usepackage{mathrsfs}
\usepackage{mathtools}
\usepackage{amsfonts}
\usepackage{textcomp}
\usepackage{multicol}
\usepackage{multirow}
\usepackage{color}
\usepackage{fixltx2e}
\usepackage{latexsym}
\usepackage{a4wide}
\usepackage{multirow}
\usepackage{float}
\usepackage{url}
\usepackage{booktabs}   

\usepackage{hyperref}
\usepackage{subcaption}
\usepackage{schemabloc,tikz,amsthm}
\usepackage{subfig}
\usepackage{tabularx}
\usepackage[numbers,sort&compress]{natbib}
\usetikzlibrary{circuits, arrows}
\usepackage{lipsum}
\usepackage[letterpaper, left=0.625in, right=0.625in, top=0.75in, bottom=.75in, footskip=0.25in]{geometry}
\usepackage{xcolor}

\usepackage{tikz}
\usetikzlibrary{plotmarks}
\usepackage{pgfplots}
\usetikzlibrary{spy}
\usepackage{collectbox}
\usepgfplotslibrary{groupplots}
\usetikzlibrary{calc}
\usetikzlibrary{spy}
\usepackage{pgfplots}
 \usepgfplotslibrary{groupplots}
\usepackage{collectbox}
\usetikzlibrary{fit,backgrounds}
\pgfplotsset{compat=1.18} 
\usepackage{soul,color}
\usepackage{subcaption}

\theoremstyle{remark}
\newtheorem{remark}{Remark}
\tikzset{new spy style/.style={spy scope={%
			magnification=2.8,
			size=1.25cm,
			connect spies,
			every spy on node/.style={
				rectangle,
				draw,
			},
			every spy in node/.style={
				draw,
				rectangle,
			}
		}
	}
}

\usetikzlibrary{arrows.meta,positioning,fit,calc}
\definecolor{splitgreen}{HTML}{A9E494}   
\definecolor{palegreen}{HTML}{E4F8EE}    
\definecolor{encblue}{HTML}{8CD2E4}      
\definecolor{modyellow}{HTML}{FDF431}    
\definecolor{concatviolet}{HTML}{DB94E4} 
\definecolor{demodyellow}{HTML}{FADD1B}  
\definecolor{sicpink}{HTML}{E494A2}      
\definecolor{detblue}{HTML}{8DDAFC}      
\definecolor{zoomblue}{HTML}{1E29F8}     
\definecolor{zoomred}{HTML}{CF2411}      
\definecolor{splitgreen}{HTML}{A9E494}   
\definecolor{palegreen}{HTML}{E4F8EE}    
\definecolor{encblue}{HTML}{8CD2E4}      
\definecolor{modyellow}{HTML}{FDF431}    
\definecolor{concatviolet}{HTML}{DB94E4} 
\definecolor{demodyellow}{HTML}{FADD1B}  
\definecolor{sicpink}{HTML}{E494A2}      
\definecolor{detblue}{HTML}{8DDAFC}      
\definecolor{zoomblue}{HTML}{1E29F8}     
\definecolor{zoomred}{HTML}{CF2411}      
\definecolor{comline}{HTML}{0057D8}      
\definecolor{privline}{HTML}{007F3B}     
\pgfplotscreateplotcyclelist{mycolorlist}{%
	black,densely dashed,every mark/.append style={fill=black!80!black},mark=o\\
	brown,every mark/.append style={fill=brown!80!black},mark=otimes\\%
	red,every mark/.append style={fill=red!80!black},mark=otimes\\%
	blue,every mark/.append style={fill=blue!80!black},mark=otimes\\%
	orange,every mark/.append style={fill=blue!80!black},mark=otimes\\%
	red,densely dashed,every mark/.append style={fill=black!80!black},mark=o\\
	blue,densely dashed,every mark/.append style={fill=black!80!black},mark=o\\
	brown,densely dashed,every mark/.append style={fill=black!80!black},mark=o\\
}

\begin{document}
\thispagestyle{empty}
\pagestyle{empty}
\tikzset{new spy style/.style={spy scope={%
			magnification=2.8,
			size=1.25cm,
			connect spies,
			every spy on node/.style={
				rectangle,
				draw,
			},
			every spy in node/.style={
				draw,
				rectangle,
			}
		}
	}
}
\newcolumntype{L}[1]{>{\raggedright\let\newline\\\arraybackslash\hspace{0pt}}m{#1}}
\newcolumntype{C}[1]{>{\centering\let\newline\\\arraybackslash\hspace{0pt}}m{#1}}
\newcolumntype{R}[1]{>{\raggedleft\let\newline\\\arraybackslash\hspace{0pt}}m{#1}}
\pgfdeclarelayer{background}
\pgfdeclarelayer{foreground}
\pgfsetlayers{background,main,foreground}
\newcommand{\oeq}{\mathrel{\text{\sqbox{$=$}}}}
\setlength{\textfloatsep}{0.1cm}
\setlength{\floatsep}{0.1cm}
\tikzstyle{int}=[draw, fill=white!20, minimum size=2em]
\tikzstyle{init} = [pin edge={to-,thin,black}]
\title{
\vspace{-.3em} }
\title{Error Performance Analysis of CFO-Impaired OFDM-Based Rate-Splitting SCMA}

    \author{Minerva Priyadarsini,~\IEEEmembership{Student Member,~IEEE,}%
        ~Kuntal Deka,~\IEEEmembership{Member,~IEEE,}
         ~Zilong Liu,~\IEEEmembership{Senior Member,~IEEE,}
        ~Sanjeev Sharma,~\IEEEmembership{Senior Member,~IEEE}%
        \thanks{Minerva Priyadarsini is with the School of Electrical Sciences, Indian Institute of Technology Goa, Goa 40301, India (email: minerva183212005@iitgoa.ac.in).}%
        \vspace{-.3em}
        \thanks{Kuntal Deka is with the Department of EEE,  Indian Institute of Technology Guwahati, Guwahati 780139, India (email: kuntaldeka@iitg.ac.in).}%
         \thanks{Zilong Liu is with  is with the School of Computer Science and Electrical
Engineering, University of Essex, Colchester CO4 3SQ, U.K. (e-mail:
zilong.liu@essex.ac.uk).}%
        \thanks{Sanjeev Sharma is with the Department of Electronics Engineering,  Indian Institute of Technology (BHU), India (email: sanjeev.ece@iitbhu.ac.in).}%
}

\maketitle
\thispagestyle{empty} 

\begin{abstract}
Rate-splitting sparse code multiple access (RS-SCMA) superimposes
QAM-modulated common messages and SCMA-encoded private messages
over the same resource elements. This paper analyzes the bit error rate (BER) performance of orthogonal frequency division multiplexing (OFDM)-based RS-SCMA in the presence of carrier frequency offset (CFO) over frequency-selective Rayleigh fading channels. A finite-alphabet BER framework is developed for common-layer detection, successive interference cancellation (SIC), and private-layer SCMA detection. The proposed analysis explicitly retains the common and private-layer signal components and models the unresolved intra- and inter-block CFO leakage as conditionally Gaussian interference using its conditional second-order statistics. Residual interference caused by erroneous common-layer cancellation and the effect of a tunable message-splitting factor are also incorporated. The results show that CFO produces a high-SNR BER floor that increases with the normalized frequency offset. Under the
considered conditions, the overloaded SCMA private layer is more
severely affected and dominates the overall BER. Simulations with perfect SIC further show that eliminating common-layer decision errors has only a marginal impact on the private-layer BER floor, indicating that the dominant degradation arises from CFO-impaired private-layer detection rather than SIC error propagation. Monte Carlo simulations validate the derived BER expressions and demonstrate their accuracy in capturing the layer-wise and overall BER performance over the considered CFO range.
\end{abstract}


\section{Introduction}

\subsection{Background and Related Works}

Next-generation wireless networks require multiple-access schemes
that support massive connectivity and high spectral efficiency while
maintaining reliable transmission under multiuser interference (MUI)
\cite{6g_survey,V2X,6g_new1}. These requirements have motivated
non-orthogonal multiple-access techniques that allow users to share
time, frequency, and code resources. Sparse code multiple access
(SCMA) is a code-domain non-orthogonal multiple-access (NOMA) scheme
that maps user data onto sparse multidimensional codewords
\cite{SCMA1,scma_cb}. This mapping supports overloaded transmission
over shared resource elements (RE). At the receiver, the message-passing
algorithm (MPA) exploits the sparse factor graph to perform multiuser
detection with manageable computational complexity. 
Over the years, considerable effort has been devoted to characterizing the BER performance of SCMA systems under different channel conditions. For star-QAM-based SCMA codebooks, the BER was analyzed over additive white Gaussian noise (AWGN) channels in \cite{Yu_letter}. This analysis was subsequently extended in \cite{Yu2018}, where an improved codebook design and exact and approximate BER expressions were developed for Rayleigh fading channels. The BER performance of CFO-impaired QAM-based orthogonal frequency-division multiplexing (OFDM) over frequency-selective fading channels was investigated in \cite{ofdm_ber}. In that work, the CFO-induced inter-carrier interference (ICI) was modeled using a Gaussian approximation, and the resulting BER expressions showed good agreement with simulation results. Since OFDM-based SCMA inherits the sensitivity of OFDM to CFO, the resulting loss of subcarrier orthogonality introduces ICI into the SCMA detection process. Building upon these developments, the BER performance of OFDM-SCMA under CFO was investigated in \cite{ofdm_scma} over both AWGN and Rayleigh fading channels.

Rate-splitting multiple access (RSMA) provides another approach to
managing MUI \cite{Dizdar2020,rsma_primer}.  RSMA divides each user's  message into common and private parts, enabling a portion of the interference to be decoded and removed through successive interference cancellation (SIC), while the remaining interference is treated as noise. Multicarrier RSMA has been investigated through joint power and subcarrier allocation \cite{mc_rsma_ofdm,mc_rsma}.    OFDM-RSMA has been studied over doubly selective channels and in multi-numerology configurations \cite{rsma_ofdm}. A DFT-based flexible RSMA framework was proposed in \cite{dft_rsma}, in which common and private streams are separated across transform domains to mitigate inter-stream interference and reduce reliance on SIC. These studies demonstrate the applicability of RSMA to wideband transmission, but primarily focus on achievable-rate optimization, power-domain interference management, and resource allocation. From an error-performance perspective, Vu \textit{et al.} \cite{err_rsma} derived exact and asymptotic symbol error rate (SER) expressions for downlink multiple-input single-output (MISO)-RSMA with  BPSK and QPSK configurations for the common and private streams, while accounting for SIC error propagation. The analysis characterized the diversity order and modulation gain of the two streams, and a min-max power-allocation scheme was proposed to improve SER fairness among users. In \cite{err_rsma_ser}, a constellation-based framework was developed for multiuser RSMA and its variants, yielding exact SER expressions that incorporate SIC error propagation. The study further proposed a coordinate-interleaved RSMA scheme to mitigate private-stream interference and derived power-allocation bounds for unambiguous symbol detection.

 The complementary strengths of RSMA and SCMA naturally lead to a broader
question: 
\textit{Instead of further growing the knowledge tree of multiple access, how to shrink and unify, so as to exploit all five dimensions (time, frequency, space, power, and signal) in the simplest way?}
This question underpins the concept of \emph{universal
multiple access} (UMA) introduced in \cite{UMA}. Integrating rate splitting
with code-domain multiple access offers a natural path toward UMA.
Following this direction, rate-splitting sparse code multiple access (RS-SCMA)
was recently proposed in \cite{RS_SCMA} as an initial step toward UMA. 
RS-SCMA is not merely a straightforward combination of RSMA and SCMA. It adopts a
hierarchical transmission structure in which an $M$-ary QAM modulated common layer
is superimposed on SCMA-encoded private streams. 
At the receiver, the common layer is decoded first and removed through SIC,
after which the private streams are recovered using MPA. MUI
is therefore handled through two complementary mechanisms: a portion is
decoded through the common layer and subsequently cancelled by SIC, while the
remaining interference is resolved through code-domain separation in the
private SCMA layer.

An important feature of RS-SCMA is the message-splitting factor, which
determines the relative lengths of the common and private message portions and,
consequently, the transmission structure. By
selecting this factor, the system can trade off error-rate performance,
achievable rate, and effective overloading according to the operating
conditions \cite{RS_SCMA}. Moreover, by integrating rate splitting with sparse
code-domain transmission, RS-SCMA exploits the interference-management.
It therefore provides a unified architecture in which message splitting and
SCMA overloading can be jointly configured, making RS-SCMA a potential
candidate for UMA.
\subsection{Motivation and Contributions}

The hierarchical RS-SCMA receiver introduces a coupling between SIC-based
common-layer detection and MPA-based private-layer detection that arises in
neither conventional SCMA nor conventional RSMA. Specifically, the common layer
is detected in the presence of the superimposed private SCMA signal, and an
erroneous common-layer decision leaves a residual after SIC that degrades the
subsequent MPA-based private-layer detection. This coupling is further
affected by CFO. In high-mobility scenarios, CFO
arises mainly from Doppler shifts, and it may also result from oscillator
mismatch between the transmitter and the receiver \cite{cfo_paper}. In OFDM
systems, CFO destroys subcarrier orthogonality and causes ICI, which, in OFDM-based RS-SCMA, affects both the common and
private layers. Consequently, the private-layer BER depends jointly on the
CFO-induced interference and on the reliability of the preceding common-layer
decision.

Existing BER analyses of OFDM-SCMA, e.g., \cite{ofdm_scma}, model the aggregate
CFO-induced interference as Gaussian. Such a model does not directly extend to
OFDM-based RS-SCMA, in which finite-alphabet common and private components are
detected successively and coupled through SIC. Meanwhile, existing
multicarrier RSMA studies \cite{mc_rsma_ofdm,rsma_ofdm,mc_rsma,dft_rsma}
consider neither sparse finite-alphabet private streams nor MPA-based
detection. These limitations call for an error analysis that preserves the
relevant finite-alphabet structure while accounting for both CFO-induced ICI
and common-to-private SIC error propagation.

To this end, this paper develops a finite-alphabet BER analysis for OFDM-based
RS-SCMA over frequency-selective Rayleigh fading channels in the presence of
CFO. The analysis retains the desired common and private finite-alphabet
components explicitly and models only the unresolved CFO-induced leakage as
conditionally Gaussian interference. The main contributions of this paper are summarized
as follows.

\begin{itemize}

\item
We develop a finite-alphabet BER analysis for CFO-impaired OFDM-based RS-SCMA
that  follows the receiver detection chain comprising common-layer $M$-QAM
detection, SIC, and MPA-based private-layer SCMA
detection. In the common-layer analysis, the finite-alphabet private SCMA
component is retained explicitly, while the private-layer analysis incorporates
the residual interference produced by erroneous common-layer cancellation.

\item
We obtain a tractable representation of the CFO-induced interference by
modeling the unresolved intra-block and out-of-block leakage through its
conditional second-order statistics. The resulting framework captures both
CFO-induced ICI and common-to-private SIC error propagation, and it applies to
any value of the tunable message-splitting factor $\alpha$.

\item
Using the developed framework, we characterize the BER of the QAM-modulated
common layer, the SCMA-encoded private layer, and the overall system for
various normalized CFO values. The results show that CFO induces a high-SNR
error floor that rises with the normalized CFO and is dominated by the private
layer. Moreover, simulations with correct-decision SIC  show that the private-layer error floor remains nearly
unchanged when common-to-private error propagation is removed. This indicates
that CFO-impaired SCMA private-layer detection, rather than SIC error propagation,
is the main source of the error floor. The analytical results are validated
through Monte Carlo simulations with MPA-based private-layer detection.

\end{itemize}

 \subsection{Notations}
In this paper, regular, bold lowercase, bold uppercase, and script fonts denote scalars, vectors, matrices, and sets, respectively. $\mathbb{C}^{J\times 1}$ represents a complex vector of dimension $J\times 1$. $\mathcal{CN}(\mu, \sigma^2)$ denotes a complex Gaussian distribution with mean $\mu$ and variance $\sigma^2$. The operators $(\cdot)^H$ and $(\cdot)^T$ represent the conjugate transpose and transpose operations, respectively. $\mathbf{I}_K$ denotes a $K \times K$ identity matrix. 
Table~\ref{tab:notation} presents various  notations and their meaning.
\begin{table}[tb]
\centering
\caption{Symbols and notations.}
\label{tab:notation}
\setlength{\tabcolsep}{2pt}
\renewcommand{\arraystretch}{0.95}
\footnotesize
\begin{tabular}{p{0.20\columnwidth}p{0.76\columnwidth}}
\hline
\textbf{Symbol} & \textbf{Meaning}\\
\hline

$J,K$
& Numbers of users and SCMA resource elements (REs), respectively.\\

$j,k$
& Receiving-user and RE indices, respectively.\\

$q,q'$
& RS-SCMA block indices within an OFDM block.\\

$Q$
& Number of RS-SCMA blocks mapped within one OFDM block.\\
$\mathrm{Q}(\cdot)$
& Gaussian $Q$-function.\\
$N_{\mathrm{sc}},\,N_{\mathrm{cp}}$
& Numbers of OFDM subcarriers and cyclic-prefix samples, respectively.\\

$\alpha$
& Message-splitting factor.\\

$l_{\mathrm c},\,l_{\mathrm p}$
& Lengths of the common and private message portions,
$l_{\mathrm c}=\alpha N$ and
$l_{\mathrm p}=(1-\alpha)N$.\\

$\lambda_{\mathrm{RS\text{-}SCMA}}$
& Effective overloading factor of RS-SCMA.\\

$M$
& Modulation order of the common layer and codebook size of the private SCMA layer.\\

$\mathbf F$
& SCMA factor-graph matrix.\\

$\mathbf x_{\mathrm c}^{(q)}$,
$\mathbf x_{\mathrm p}^{(q)}$
& Common-layer and aggregate private-layer vectors of the $q$th RS-SCMA block.\\

$\nu$
& Transmission mode,
$\nu\in\{\mathrm{JP},\mathrm P,\mathrm C\}$ for joint,
private-only, and common-only transmission, respectively.\\

$\mathbf x^{(q,\nu)}$
& Transmitted $q$th RS-SCMA block in mode $\nu$.\\

$\mathbf E_q$
& Subcarrier-insertion matrix for the $q$th RS-SCMA block.\\

$\mathbf E_{\mathrm{out}}$
& Concatenated insertion matrices of all RS-SCMA blocks other than the
desired block.\\

$\mathbf x_{\mathrm F}^{(\nu)}$
& Frequency-domain OFDM transmit vector.\\

$\mathbf x_{\mathrm T}^{(\nu)}$
& Time-domain OFDM transmit vector obtained by the
$N_{\mathrm{sc}}$-point IFFT.\\

$\mathbf U$
& Unitary $N_{\mathrm{sc}}$-point DFT matrix.\\

$\varepsilon_j$
& CFO of user $j$, normalized by the OFDM subcarrier spacing.\\

$\boldsymbol{\Phi}_j$
& Frequency-domain CFO coupling matrix of user $j$.\\

$\mathbf g_j$
& Frequency-domain channel vector of user $j$.\\

$\mathbf G_j$
& Diagonal frequency-domain channel matrix,
$\mathbf G_j=\operatorname{diag}(\mathbf g_j)$.\\

$\mathbf B_j^{(q)}$
& CFO-impaired effective channel matrix of the $q$th demapped RS-SCMA block.\\

$\mathbf H_j^{(q)}$
& Diagonal part of $\mathbf B_j^{(q)}$ used for detection.\\

$\boldsymbol{\Delta}_j^{(q)}$
& Off-diagonal part of $\mathbf B_j^{(q)}$ representing intra-block CFO leakage.\\

$\mathbf i_{\mathrm{out},j}^{(q,\nu)}$
& CFO-induced leakage from the remaining RS-SCMA blocks into the $q$th block.\\

$\mathbf u,\mathbf v$
& Common-symbol and private SCMA codeword-index vectors used in the error analysis.\\

$\mathbf x_{\mathrm c,\mathbf u}$,
$\mathbf x_{\mathrm p,\mathbf v}$
& Common-layer vector and aggregate private-layer vector associated with
$\mathbf u$ and $\mathbf v$.\\

$\mathbf A_{\mathrm{out}}(\mathbf g)$
& CFO coupling matrix from the remaining RS-SCMA blocks to the desired block.\\

$E_{\mathrm{RE}}^{(\nu)}$
& Average transmitted energy per RE in mode $\nu$.\\

$\mathbf R_I^{(\nu)}(\mathbf g)$,
$\mathbf R_{\Delta}^{(\nu)}(\mathbf g)$
& Conditional out-of-block and intra-block CFO leakage covariance matrices.\\

$\mathbf z^{(\nu)}$,
$\mathbf R_{\mathrm{eff}}^{(\nu)}(\mathbf g)$
& Effective disturbance vector and its conditional covariance matrix.\\

$\mathbf d_{\mathrm c}^{(\nu)}$,
$\mathbf d_{\mathrm p}^{(\nu)}$
& Common- and private-layer pairwise difference vectors.\\

$\mathbf r_{\mathrm{sic}}^{(\mathrm{JP})}$
& SIC residual caused by an erroneous common-layer decision.\\

$P_{\ell}^{(\nu)}$
& Conditional pairwise error probability (PEP) of layer
$\ell\in\{\mathrm c,\mathrm p\}$.\\

$B_{\mathrm c},\,B_{\mathrm p}$
& Numbers of bits conveyed by one common-layer vector and one aggregate
private-layer vector, respectively.\\

$\mathrm{BER}_{\ell}^{(\nu)}$
& Layer-wise BER for layer
$\ell\in\{\mathrm c,\mathrm p\}$ in mode $\nu$.\\

$\mathrm{BER}_{\mathrm o}$
& Overall BER of the OFDM-RS-SCMA system.\\
\hline
\end{tabular}
\end{table}
 \subsection{Organization}
The paper is organized as follows. Section~\ref{sec:prel} introduces the basic concepts of SCMA. Section~\ref{sec:system_model} presents the proposed CFO-impaired RS-SCMA system model. Section~\ref{sec:error_performance_analysis} presents the error analysis of the CFO-impaired error analysis. Section~\ref{sec:results} discusses and presents the simulation results and Section~\ref{sec:conclusion} concludes the paper.

\section{A Brief Introduction to SCMA}
\label{sec:prel}

SCMA is a code-domain NOMA scheme in which multiple users share a limited
number of orthogonal subcarriers or RE by transmitting multidimensional
sparse codewords rather than scalar modulation symbols \cite{SCMA1}.
Specifically, each user maps $\log_2 M$ bits directly to one of $M$ sparse
$K$-dimensional codewords from its own codebook, where $K$ denotes the number
of REs. When $J$ users are multiplexed over the $K$ REs with $J > K$, SCMA
supports overloaded transmission with an overloading factor of
$\lambda = J/K > 1$.

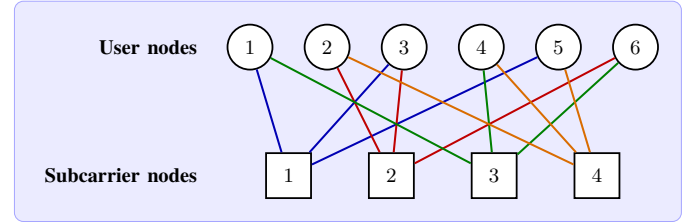
\begin{figure}[!htb]
	\begin{center}
		\scalebox{0.85}{		
\begin{tikzpicture}[
    user/.style = {circle, draw, thick, fill=white,
                   minimum size=7mm, inner sep=0pt, font=\small},
    re/.style   = {rectangle, draw, thick, fill=white,
                   minimum size=7mm, inner sep=0pt, font=\small},
    lbl/.style  = {font=\small\bfseries, anchor=east, align=right},
    edge/.style = {line width=0.9pt},
]

\foreach \j in {1,...,6}
  \node[user] (u\j) at ({1.2*(\j-1)}, 0) {$\j$};

\foreach \k in {1,...,4}
  \node[re] (r\k) at ({0.6 + 1.6*(\k-1)}, -2) {$\k$};

\node[lbl] (ulbl) at (-0.7, 0)  {User nodes};
\node[lbl] (rlbl) at (-0.7, -2) {Subcarrier nodes};

\foreach \j/\k/\c in {
    1/1/blue!70!black,   3/1/blue!70!black,   5/1/blue!70!black,
    2/2/red!75!black,    3/2/red!75!black,    6/2/red!75!black,
    1/3/green!50!black,  4/3/green!50!black,  6/3/green!50!black,
    2/4/orange!85!black, 4/4/orange!85!black, 5/4/orange!85!black}
  \draw[edge, draw=\c] (u\j) -- (r\k);

\begin{scope}[on background layer]
  \node[fit=(ulbl)(rlbl)(u1)(u6)(r1)(r4), inner sep=3.5mm,
        rounded corners=4pt, fill=blue!8, draw=blue!35] {};
\end{scope}

\end{tikzpicture}%
}
\end{center}
	\caption{\footnotesize{Factor graph of $J=6$ users and $K=4$ subcarrier or resource nodes with $d_v=2$ and $d_f=3$ corresponding to the indicator matrix ${\bf{F}}$ in (\ref{F_mat}).}}
	\label{factor_graph1}
\end{figure}

The sparse mapping between users and REs is specified by an indicator matrix
\(\mathbf F\in\{0,1\}^{K\times J}\) \cite{scma_tut}, whose \((k,j)\)th entry
\(f_{k,j}\) equals 1 if the codeword of user \(j\) has a nonzero component on
RE \(k\), and 0 otherwise. Hence, the \(j\)th column of \(\mathbf F\)
indicates the REs occupied by user \(j\), and the \(k\)th row indicates the
users superimposed on RE \(k\). Let \(d_v\) and \(d_f\) denote the number of
nonzero entries in each user codeword and the number of users colliding on
each RE, respectively, i.e., the column and row weights of \(\mathbf F\). For
example, for \(J=6\), \(K=4\), \(d_v=2\), and \(d_f=3\), a possible indicator
matrix is
\begin{equation}
\mathbf F=
\begin{bmatrix}
1&0&1&0&1&0\\
0&1&1&0&0&1\\
1&0&0&1&0&1\\
0&1&0&1&1&0
\end{bmatrix}.
\label{F_mat}
\end{equation}
The corresponding factor graph is shown in Fig.~\ref{factor_graph1}, where
circles and squares represent user nodes and RE nodes, respectively. The
structure of this graph specifies which users collide on each RE, and its
sparsity enables low-complexity detection via  MPA.

Each user \(j\) is assigned a dedicated codebook \cite{scma_cb,Li2022,Deka_DE}
\begin{equation}
    \mathcal C_j
    =
    \left[
    \mathbf c_{j,1},
    \mathbf c_{j,2},
    \ldots,
    \mathbf c_{j,M}
    \right]
    \in\mathbb C^{K\times M},
    \label{eq:scma_codebook_prelim}
\end{equation}
where each codeword \(\mathbf c_{j,m}\in\mathbb C^{K\times1}\) has \(d_v\)
nonzero entries at the RE positions indicated by the \(j\)th column of
\(\mathbf F\). User \(j\) maps \(\log_2 M\) bits to an index
\(m\in\{1,\ldots,M\}\) and transmits the corresponding codeword
\(\mathbf c_{j,m}\). The codewords of all \(J\) users are superimposed over
the same \(K\) REs.

At the receiver, multiuser detection is performed  using the MPA
\cite{SCMA_det_low_complx,SCMA_det_low_complx1}, in which likelihood messages
are exchanged iteratively between user nodes and RE nodes along the edges of
the factor graph. Since only \(d_f\) users collide on each RE, each RE node
processes \(M^{d_f}\) codeword combinations rather than the \(M^J\)
combinations required by exhaustive maximum-likelihood (ML) detection
\cite{Wu2015}.

\section{CFO-Impaired OFDM-RS-SCMA System Model}
\label{sec:system_model}
\begin{figure*}[!htbp]
\centering
\resizebox{\textwidth}{!}{%
\begin{tikzpicture}[
  >=Stealth, line width=0.7pt,
  font=\small,
  blk/.style={draw, align=center, fill=white, inner sep=3pt},
  big/.style={blk, text width=2.6cm, minimum height=1.5cm},
  chn/.style={blk, inner sep=2pt},
  lbl/.style={font=\small, inner sep=1.5pt},
  arr/.style={->},
  carr/.style={->, comline},
  parr/.style={->, privline},
  zoombox/.style={draw=zoomblue, dashed, thick},
  zoom/.style={zoomred, thick, dash pattern=on 6pt off 2pt on 1pt off 2pt},
]

\node[big] (tx1) at ( 2.6,-2.8) {RS-SCMA\\Transmission\\(Block $1$)};
\node[big] (txq) at ( 2.6,-5.1) {RS-SCMA\\Transmission\\(Block $q$)};
\node[big] (txQ) at ( 2.6,-7.4) {RS-SCMA\\Transmission\\(Block $Q$)};

\node[chn, minimum height=2.6cm]  (alloc)   at ( 7.30,-4.2) {Subcarrier\\Mapping\\[1mm]$\{\mathbf E_q\}_{q=1}^{Q}$};
\node[chn, minimum height=1.7cm]  (ifft)    at ( 9.75,-4.2) {IDFT\\[1mm]$\mathbf U^{\mathrm H}$};
\node[chn, minimum height=1.45cm] (addcp)   at (11.75,-4.2) {Add\\CP};
\node[chn, minimum height=1.0cm]  (ps)      at (13.05,-4.2) {P/S};
\node[chn, minimum height=2.4cm]  (chan)    at (14.85,-4.2) {Channel\\$\mathbf h_j$\\[1mm]CFO\\$\varepsilon_j$};
\node[chn, minimum height=1.0cm]  (sp)      at (16.65,-4.2) {S/P};
\node[chn, minimum height=1.5cm]  (rmcp)    at (17.95,-4.2) {Remove\\CP};
\node[chn, minimum height=1.7cm]  (fft)     at (19.95,-4.2) {DFT\\[1mm]$\mathbf U$};
\node[chn, minimum height=2.6cm]  (dealloc) at (22.30,-4.2) {Subcarrier\\Demapping\\[1mm]$\{\mathbf E_q^{\mathrm H}\}_{q=1}^{Q}$};

\node[big] (det1) at (26.4,-2.8) {RS-SCMA\\Detection\\(Block $1$)};
\node[big] (detq) at (26.4,-5.1) {RS-SCMA\\Detection\\(Block $q$)};
\node[big] (detQ) at (26.4,-7.4) {RS-SCMA\\Detection\\(Block $Q$)};

\draw[arr] (tx1.east) node[lbl,above right=1pt and 2pt]{$\mathbf{x}^{(1,\nu)}$}
           -- ++(0.9,0) |- ([yshift= 0.85cm]alloc.west);
\draw[arr] (txq.east) node[lbl,above right=1pt and 2pt]{$\mathbf{x}^{(q,\nu)}$}
           -- ++(0.9,0) |- ([yshift= 0.00cm]alloc.west);
\draw[arr] (txQ.east) node[lbl,above right=1pt and 2pt]{$\mathbf{x}^{(Q,\nu)}$}
           -- ++(0.9,0) |- ([yshift=-0.85cm]alloc.west);
\node at (5.25,-3.80) {$\vdots$};
\node at (5.25,-6.10) {$\vdots$};

\draw[arr] (alloc) -- node[lbl,above]{$\mathbf x_{\mathrm F}^{(\nu)}$} (ifft);
\draw[arr] (ifft)  -- node[lbl,above]{$\mathbf x_{\mathrm T}^{(\nu)}$} (addcp);
\draw[arr] (addcp) -- (ps);
\draw[arr] (ps)    -- (chan);
\draw[arr] (chan)  -- (sp);
\draw[arr] (sp)    -- (rmcp);
\draw[arr] (rmcp)  -- node[lbl,above]{$\mathbf y_{\mathrm T,j}^{(\nu)}$} (fft);
\draw[arr] (fft)   -- node[lbl,above]{$\mathbf y_{\mathrm F,j}^{(\nu)}$} (dealloc);

\node[lbl] (awgn) at (14.85,-2.25) {$\mathbf w_{\mathrm T,j}$};
\draw[arr] (awgn) -- (chan.north);

\draw[arr] ([yshift= 0.85cm]dealloc.east) -- ++(0.6,0) |- (det1.west)
           node[lbl, above left=1pt and 1pt]{$\mathbf{y}_j^{(1,\nu)}$};
\draw[arr] ([yshift= 0.00cm]dealloc.east) -- ++(0.6,0) |- (detq.west)
           node[lbl, above left=1pt and 1pt]{$\mathbf{y}_j^{(q,\nu)}$};
\draw[arr] ([yshift=-0.85cm]dealloc.east) -- ++(0.6,0) |- (detQ.west)
           node[lbl, above left=1pt and 1pt]{$\mathbf{y}_j^{(Q,\nu)}$};
\node at (24.15,-3.80) {$\vdots$};
\node at (24.15,-6.10) {$\vdots$};

\node[lbl, font=\small\bfseries] at (2.6,-1.35) {BS};
\node[lbl, font=\small\bfseries] at (26.4,-1.35) {User $j$};

\draw[zoombox] (1.4,-18.0) rectangle (14.6,-8.4);
\node[lbl, anchor=west] at (1.7,-17.5) {RS-SCMA Transmitter (Block $q$)};

\node[blk, fill=splitgreen, minimum width=1.5cm, minimum height=0.55cm] (s1)  at (3.65,-10.50) {Split};
\node[blk, fill=splitgreen, minimum width=1.5cm, minimum height=0.55cm] (s2)  at (3.65,-11.40) {Split};
\node[blk, fill=splitgreen, minimum width=1.5cm, minimum height=0.55cm] (sK)  at (3.65,-12.75) {Split};
\node at (3.65,-11.95) {$\vdots$};
\node at (3.65,-15.65) {$\vdots$};

\draw[arr] (1.75,-10.50) node[lbl,above right=1pt and -2pt]{$\mathbf{b}_1$} -- (s1.west);
\draw[arr] (1.75,-11.40) node[lbl,above right=1pt and -2pt]{$\mathbf{b}_2$} -- (s2.west);
\draw[arr] (1.75,-12.75) node[lbl,above right=1pt and -2pt]{$\mathbf{b}_K$} -- (sK.west);

\node[blk, fill=concatviolet, minimum height=3.1cm]  (concat) at ( 7.00,-10.20) {Concatenate\\Common\\Parts};
\node[blk, fill=modyellow,    minimum height=3.05cm] (mod)    at ( 9.85,-10.20) {$M$-QAM\\Modulator};
\node[blk, fill=encblue,      minimum height=4.9cm]  (enc)    at ( 9.85,-14.35) {SCMA\\Encoder\\[1mm]$\{\mathcal C_j\}_{j=1}^{J}$};
\node[blk, fill=splitgreen, minimum width=1.05cm, minimum height=8.2cm] (sup)    at (12.70,-12.85) {};
\node[rotate=90, align=center] at (sup) {Power Scaling and Superposition};

\draw (s1.east) -- ++(0.25,0) coordinate (f1);
\draw[carr] (f1) |- node[lbl,pos=0.75,above,text=comline]{$\mathbf{b}_1^{\mathrm{c}}$} ([yshift= 1.05cm]concat.west);
\draw[parr] (f1) |- node[lbl,pos=0.75,above,text=privline]{$\mathbf{b}_1^{\mathrm{p}}$} ([yshift= 2.00cm]enc.west);
\draw (s2.east) -- ++(0.50,0) coordinate (f2);
\draw[carr] (f2) |- node[lbl,pos=0.78,above,text=comline]{$\mathbf{b}_2^{\mathrm{c}}$} ([yshift= 0.35cm]concat.west);
\draw[parr] (f2) |- node[lbl,pos=0.75,above,text=privline]{$\mathbf{b}_2^{\mathrm{p}}$} ([yshift= 1.30cm]enc.west);
\draw (sK.east) -- ++(0.75,0) coordinate (fK);
\draw[carr] (fK) |- node[lbl,pos=0.80,above,text=comline]{$\mathbf{b}_K^{\mathrm{c}}$} ([yshift=-0.80cm]concat.west);
\draw[parr] (fK) |- node[lbl,pos=0.80,above,text=privline]{$\mathbf{b}_K^{\mathrm{p}}$} ([yshift= 0.55cm]enc.west);
\draw[parr] (1.75,-15.10) node[lbl,above right=1pt and -2pt,text=black]{$\mathbf{b}_{K+1}$} -- (1.75,-15.10 -| enc.west);
\draw[parr] (1.75,-16.45) node[lbl,above right=1pt and -2pt,text=black]{$\mathbf{b}_{J}$}   -- (1.75,-16.45 -| enc.west);

\draw[arr] (concat.east) -- node[lbl,above]{$\mathbf{b}_{\mathrm{c}}$} (mod.west);
\draw[arr] (mod.east)    -- node[lbl,above]{$\mathbf{x}_{\mathrm{c}}^{(q)}$} (mod.east -| sup.west);

\draw[arr] ([yshift= 2.20cm]enc.east) -- node[lbl,above]{$\mathbf{x}_1^{(q)}$}     ([yshift= 2.20cm]enc.east -| sup.west);
\draw[arr] ([yshift= 1.55cm]enc.east) -- node[lbl,above]{$\mathbf{x}_2^{(q)}$}     ([yshift= 1.55cm]enc.east -| sup.west);
\draw[arr] ([yshift= 0.40cm]enc.east) -- node[lbl,above]{$\mathbf{x}_K^{(q)}$}     ([yshift= 0.40cm]enc.east -| sup.west);
\draw[arr] ([yshift=-0.40cm]enc.east) -- node[lbl,above]{$\mathbf{x}_{K+1}^{(q)}$} ([yshift=-0.40cm]enc.east -| sup.west);
\draw[arr] ([yshift=-1.80cm]enc.east) -- node[lbl,above]{$\mathbf{x}_{J}^{(q)}$}   ([yshift=-1.80cm]enc.east -| sup.west);
\node at (11.95,-13.35) {$\vdots$};
\node at (11.95,-15.65) {$\vdots$};

\draw[arr] (sup.east) -- node[lbl,above]{$\mathbf{x}^{(q,\nu)}$} ++(1.35,0);

\node[lbl, anchor=east] at (14.3,-17.5)
  {$\mathbf x^{(q,\nu)}=\sqrt{p_{\mathrm c}^{(\nu)}}\,\mathbf x_{\mathrm c}^{(q)}
   +\sqrt{p_{\mathrm p}^{(\nu)}}\sum_{i=1}^{J}\mathbf x_i^{(q)}\in\mathbb C^{K\times 1}$};

\draw[zoombox] (16.5,-16.4) rectangle (27.2,-10.0);
\node[lbl, anchor=west] at (16.8,-16.0) {RS-SCMA Receiver at User $j$ (Block $q$)};

\node[blk, fill=demodyellow, minimum height=1.85cm] (dem)     at (19.35,-11.30) {Common-Layer\\Detection ($M$-QAM)};
\node[blk, fill=sicpink, minimum width=1.4cm, minimum height=0.85cm]  (sic)     at (19.35,-13.75) {SIC};
\node[blk, fill=detblue, minimum height=3.5cm]      (scmadet) at (22.55,-13.75) {MPA-Based\\SCMA\\Detector};
\node[blk, fill=palegreen, minimum width=0.85cm, minimum height=4.65cm] (comb)  at (25.60,-13.15) {};
\node[rotate=90] at (comb) {Combiner};

\draw (16.5,-11.30) node[lbl,above right=1pt and 0pt]{$\mathbf{y}_j^{(q,\nu)}$} -- ++(0.55,0) coordinate (fy);
\draw[arr] (fy) -- (dem.west);
\draw[arr] (fy) |- (sic.west);

\draw[arr] (dem.south) -- node[lbl,right]{$\mathbf x_{\mathrm c,\widehat{\mathbf u}}$} (sic.north);

\draw[arr] (dem.east) -- node[lbl,above,pos=0.30]{$\widehat{\mathbf{b}}_{\mathrm{c}}$} (dem.east -| comb.west);

\draw[arr] (sic.east) -- node[lbl,above]{$\widetilde{\mathbf y}_j^{(q,\nu)}$} (sic.east -| scmadet.west);

\draw[arr] ([yshift= 1.60cm]scmadet.east) -- node[lbl,above]{$\widehat{\mathbf{b}}_1^{\mathrm p}$} ([yshift= 1.60cm]scmadet.east -| comb.west);
\draw[arr] ([yshift= 0.95cm]scmadet.east) -- node[lbl,above]{$\widehat{\mathbf{b}}_2^{\mathrm p}$} ([yshift= 0.95cm]scmadet.east -| comb.west);
\draw[arr] ([yshift=-0.85cm]scmadet.east) -- node[lbl,above]{$\widehat{\mathbf{b}}_J^{\mathrm p}$} ([yshift=-0.85cm]scmadet.east -| comb.west);
\node at (24.45,-13.70) {$\vdots$};

\draw[arr] (comb.east) -- node[lbl,above]{$\widehat{\mathbf{b}}_j$} ++(0.95,0);

\draw[zoom,->] (14.60,-8.60) -- (15.55,-8.60) -- (15.55,-5.70)
               -- ($(txq.east)+(0.05,-0.60)$);
\draw[zoom,->] (27.20,-10.30) -- (28.40,-10.30) -- (28.40,-5.70)
               -- ($(detq.east)+(0.05,-0.60)$);

\end{tikzpicture}%
}
\caption{Block diagram of the  OFDM-based RS-SCMA system. The $Q$
RS-SCMA blocks of an OFDM symbol are mapped onto interleaved subcarriers, and
$\nu\in\{\mathrm{JP},\mathrm P,\mathrm C\}$ denotes the transmission mode. The
insets show the transmitter and the receiver at user $j$ for the $q$th block,
where the common layer is detected and removed through SIC before MPA-based
detection of the private layer. For $j>K$, $\mathbf b_j^{\mathrm p}=\mathbf b_j$.}
\label{fig:ofdm_rsscma_block_diagram}
\end{figure*}
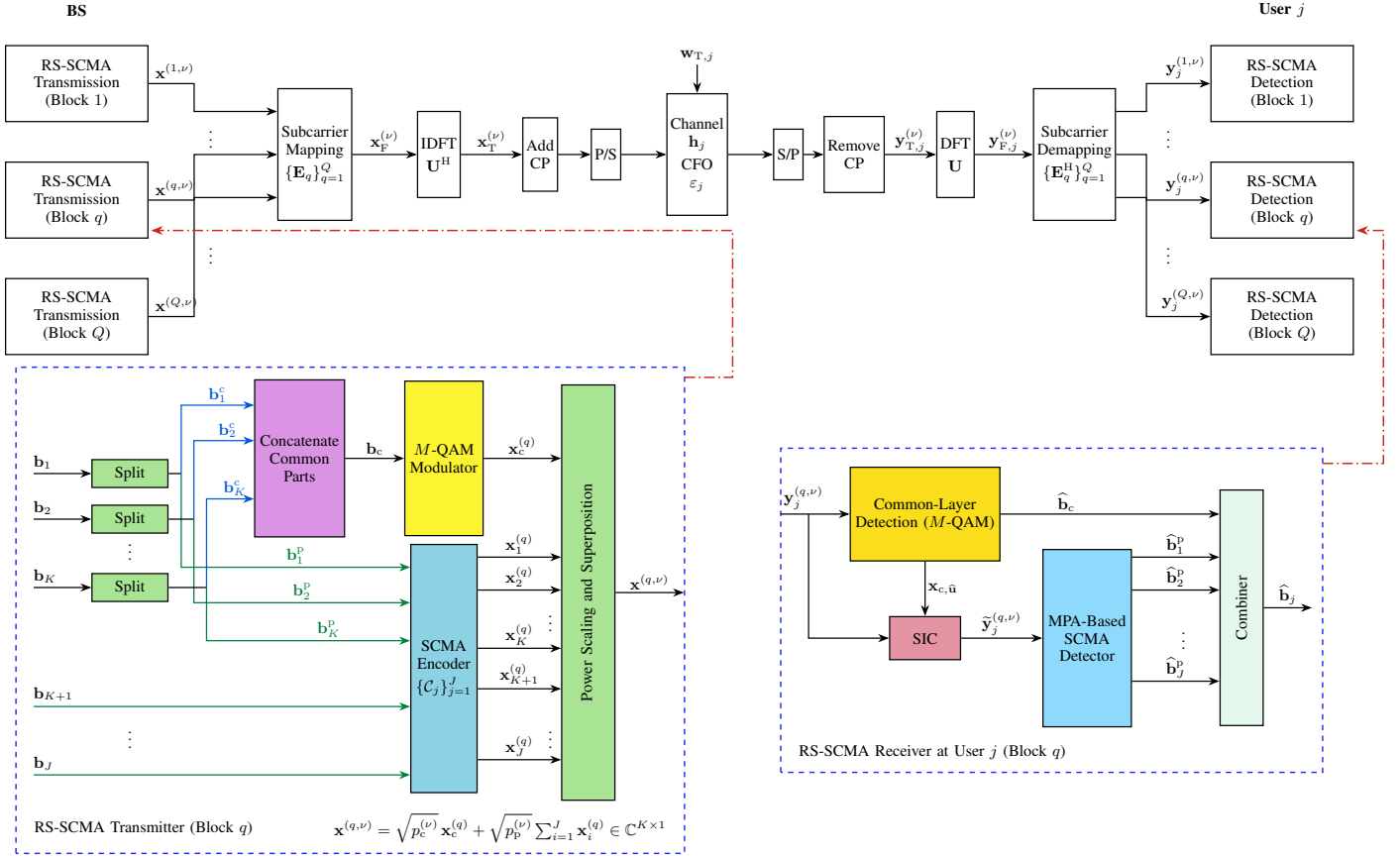
Consider the downlink OFDM-based RS-SCMA system shown in
Fig.~\ref{fig:ofdm_rsscma_block_diagram}, in which a base station (BS) serves
$J$ users over $K$ REs. Let $\mathcal J=\{1,\ldots,J\}$ and
$\mathcal K=\{1,\ldots,K\}$ denote the sets of user and RE indices,
respectively. The system adopts the  RS-SCMA architecture of
\cite{RS_SCMA}, in which an $M$-ary QAM modulated common layer is superimposed on
SCMA-encoded private streams. For completeness, we briefly review the message
splitting, signal construction, effective overloading, and detection
procedures of RS-SCMA, which form the basis of the subsequent analysis of OFDM
transmission under CFO.
\subsection{RS-SCMA Message Splitting and Signal Construction}
\label{subsec:rsscma_signal}

Let $\mathbf b_j\in\{1,\ldots,M\}^{N}$, $j\in\mathcal J$, denote the sequence
of $N$ message symbols of user $j$, and let $\alpha\in[0,1]$ denote the
message-splitting factor. For users $j\in\{1,\ldots,K\}$, the sequence
$\mathbf b_j$ is split into a common part and a private part as
\begin{equation}
    \mathbf b_j
    =
    \left[
    \left(\mathbf b_j^{\mathrm c}\right)^{\mathsf T},
    \left(\mathbf b_j^{\mathrm p}\right)^{\mathsf T}
    \right]^{\mathsf T},
    \label{eq:rsscma_split_sequences}
\end{equation}
where $\mathbf b_j^{\mathrm c}\in\{1,\ldots,M\}^{l_{\mathrm c}}$ and
$\mathbf b_j^{\mathrm p}\in\{1,\ldots,M\}^{l_{\mathrm p}}$, with
\begin{equation}
    l_{\mathrm c}=\alpha N,
    \qquad
    l_{\mathrm p}=(1-\alpha)N,
    \label{eq:rsscma_split_lengths}
\end{equation}
and $N$ and $\alpha$ are chosen such that $l_{\mathrm c}$ and $l_{\mathrm p}$
are integers. The remaining $J-K$ users contribute only to the private layer,
i.e., their entire message sequences are SCMA encoded.

In each channel use in which both common and private symbols are available,
one common symbol from each of the $K$ users is collected into
$\mathbf b_{\mathrm c}=[b_{1}^{\mathrm c},\ldots,b_{K}^{\mathrm c}]^{\mathrm T}$,
where $b_{k}^{\mathrm c}$ denotes the common symbol of user $k$ in the current
channel use; the channel-use index is omitted for notational brevity. Each
entry of $\mathbf b_{\mathrm c}$ is mapped to a point of an $M$-ary QAM
constellation $\mathcal A_{\mathrm c}$, yielding the common-layer vector
\begin{equation}
    \mathbf x_{\mathrm c}
    =
    \left[
    x_{\mathrm c,1},
    \ldots,
    x_{\mathrm c,K}
    \right]^{\mathrm T}
    \in\mathcal A_{\mathrm c}^{K},
    \label{eq:rsscma_common_vector}
\end{equation}
where $x_{\mathrm c,k}$, the modulated common symbol of user $k$, is
transmitted on RE $k$. In the same channel use, the private symbols of all
$J$ users are encoded independently using their respective SCMA codebooks.
The private symbol of user $j$ is mapped to a sparse codeword
$\mathbf x_j\in\mathcal C_j$, where
$\mathbf x_j\in\mathbb C^{K\times1}$. The aggregate private-layer vector is
then
\begin{equation}
    \mathbf x_{\mathrm p}
    =
    \sum_{j=1}^{J}\mathbf x_j.
    \label{eq:rsscma_aggregate_private}
\end{equation}

When both layers are active, the transmitted RS-SCMA signal vector is obtained
by superimposing the common-layer vector and the aggregate private-layer
vector as
\begin{equation}
    \mathbf x
    =
    \sqrt{p_{\mathrm c}}\,\mathbf x_{\mathrm c}
    +
    \sqrt{p_{\mathrm p}}\,\mathbf x_{\mathrm p},
    \label{eq:rsscma_joint_signal}
\end{equation}
where $p_{\mathrm c}$ and $p_{\mathrm p}$ denote the power allocation
coefficients of the common and private layers, respectively.

\subsubsection{Equal and Unequal Message Splitting}
The splitting factor $\alpha$ determines length of $l_{\mathrm c}$ and $l_{\mathrm p}$,  and
and their relative lengths determine the resulting transmission structure. The common and private layers are transmitted jointly over the first
\begin{equation}
    \min(l_{\mathrm c},l_{\mathrm p})
=
\min(\alpha,1-\alpha)N
\label{min_l}
\end{equation}
channel uses, as follows from \eqref{eq:rsscma_split_lengths}. This interval is
referred to as the \emph{joint phase}. If
$l_{\mathrm c}\neq l_{\mathrm p}$, the remaining
\begin{equation}
 |l_{\mathrm c}-l_{\mathrm p}|
=
|1-2\alpha|N   
\label{min_alpha}
\end{equation}
channel uses contain only the longer layer and form a subsequent
\emph{single-layer phase}. Accordingly, three cases arise depending on
$\alpha$.

For $\alpha=0.5$, the common and private portions have equal lengths,
$l_{\mathrm c}=l_{\mathrm p}=0.5N$, and only the {joint phase} is
present. For $\alpha<0.5$, the joint phase lasts for $\alpha N$ channel uses
and is followed by a {private-only phase} of length $(1-2\alpha)N$; this
case is referred to as \emph{private-dominant} transmission. For
$\alpha>0.5$, the joint phase lasts for $(1-\alpha)N$ channel uses and is
followed by a {common-only phase} of length $(2\alpha-1)N$; this case is
referred to as \emph{common-dominant} transmission.
\subsubsection{Effective Overloading Factor}

Let $K_{\mathrm c}$ and $K_{\mathrm p}$ denote the numbers of common and
private symbols transmitted per channel use when the corresponding layer is
active. For the considered RS-SCMA system,
$K_{\mathrm c}=K$ and $K_{\mathrm p}=J$. For $\alpha\neq0.5$, let
$K_{\mathrm{dom}}$ denote the number of symbols transmitted per channel use
during the single-layer phase, where
$K_{\mathrm{dom}}=K_{\mathrm p}$ for $\alpha<0.5$ and
$K_{\mathrm{dom}}=K_{\mathrm c}$ for $\alpha>0.5$.

The overloading factors of the joint and single-layer phases are
\begin{equation}
    \lambda_1
    =
    \frac{K_{\mathrm c}+K_{\mathrm p}}{K},
    \qquad
    \lambda_2
    =
    \frac{K_{\mathrm{dom}}}{K}.
    \label{eq:rsscma_phase_overloading}
\end{equation}
Since $K_{\mathrm c}=K$ and $K_{\mathrm p}=J$,
\[
    \lambda_1
    =
    \frac{K+J}{K},
\]
while
$\lambda_2=J/K$ for private-dominant transmission and
$\lambda_2=1$ for common-dominant transmission. Thus, the joint phase has a
higher loading than conventional SCMA, whereas the common-only phase is not
overloaded.

For unequal splitting, the joint and single-layer phases have different
durations and convey different numbers of symbols. The effective overloading
factor is therefore obtained by weighting the phase-wise overloading factors
by the number of symbols transmitted in each phase:
\begin{equation}
    \lambda_{\mathrm{RS\text{-}SCMA}}
    =
    \frac{w_1\lambda_1+w_2\lambda_2}{w_1+w_2},
    \label{eq:rsscma_effective_overloading}
\end{equation}
where
\begin{align}
    w_1
    &=
    \min\{l_{\mathrm c},l_{\mathrm p}\}
    (K_{\mathrm c}+K_{\mathrm p})
    \nonumber\\
    &=
    \min\{\alpha,1-\alpha\}
    (K_{\mathrm c}+K_{\mathrm p})N,
    \nonumber\\
    w_2
    &=
    |l_{\mathrm c}-l_{\mathrm p}|K_{\mathrm{dom}}
    \nonumber\\
    &=
    |1-2\alpha|K_{\mathrm{dom}}N .
    \label{eq:rsscma_effective_overloading_alpha}
\end{align}
Here, $w_1$ and $w_2$ are the numbers of symbols conveyed during the joint
and single-layer phases, respectively. Since $N$ is common to both terms, it
cancels in \eqref{eq:rsscma_effective_overloading}.
For $\alpha=0.5$, the single-layer phase is absent and $w_2=0$. Hence,
\begin{equation}
    \lambda_{\mathrm{RS\text{-}SCMA}}
    =
    \lambda_1
    =
    \frac{K+J}{K}.
\end{equation}

\textit{Example:}
For example, consider $J=6$ and $K=4$, for which $K_{\mathrm c}=4$ and
$K_{\mathrm p}=6$. At $\alpha=0.5$, only the joint phase is present, yielding
$\lambda_{\mathrm{RS\text{-}SCMA}}=\lambda_1=(4+6)/4=250\%$, compared with
$150\%$ for conventional SCMA. For $\alpha=0.25$, the joint phase with
$\lambda_1=250\%$ lasts $0.25N$ channel uses and is followed by a private-only
phase with $\lambda_2=150\%$ lasting $0.5N$ channel uses. With $w_1=2.5N$ and
$w_2=3N$, \eqref{eq:rsscma_effective_overloading} gives
$\lambda_{\mathrm{RS\text{-}SCMA}}\approx 195.45\%$.

For greater details of RS-SCMA, the reader may refer to \cite{RS_SCMA}. 

\subsection{OFDM Mapping and CFO-Impaired Frequency-Selective Channel}
\label{subsec:ofdm_cfo_model}

The $K$-dimensional RS-SCMA blocks are mapped onto the subcarriers of an OFDM
symbol for transmission over a frequency-selective channel as shown in Fig.~\ref{fig:ofdm_rsscma_block_diagram}. Each OFDM symbol
accommodates $Q$ RS-SCMA blocks, indexed by $q\in\mathcal Q=\{1,\ldots,Q\}$,
so that the total number of subcarriers is
\begin{equation}
    N_{\mathrm{sc}}=QK,
    \label{eq:nsc_qk}
\end{equation}
with subcarrier index set $\mathcal N_{\mathrm{sc}}=\{1,\ldots,N_{\mathrm{sc}}\}$.


For the $q$th block, let $\mathbf x_{\mathrm c}^{(q)}$ and
$\mathbf x_{\mathrm p}^{(q)}=\sum_{j\in\mathcal J}\mathbf x_j^{(q)}$
denote the common-layer and aggregate private-layer vectors, respectively,
where $\mathbf x_j^{(q)}\in\mathcal C_j$ is the SCMA codeword of user $j$.
Depending on $\alpha$, the $q$th block is transmitted in one of three modes,
$\nu\in\{\mathrm{JP},\mathrm P,\mathrm C\}$, corresponding to joint,
private-only, and common-only transmission. Its transmitted vector is
\begin{equation}
    \mathbf x^{(q,\nu)}
    =
    \begin{cases}
    \sqrt{p_{\mathrm c}^{(\mathrm{JP})}}\,
    \mathbf x_{\mathrm c}^{(q)}
    +
    \sqrt{p_{\mathrm p}^{(\mathrm{JP})}}\,
    \mathbf x_{\mathrm p}^{(q)},
    & \nu=\mathrm{JP},
    \\[1mm]
    \sqrt{p_{\mathrm p}^{(\mathrm P)}}\,
    \mathbf x_{\mathrm p}^{(q)},
    & \nu=\mathrm P,
    \\[1mm]
    \sqrt{p_{\mathrm c}^{(\mathrm C)}}\,
    \mathbf x_{\mathrm c}^{(q)},
    & \nu=\mathrm C.
    \end{cases}
    \label{eq:interval_transmit_vectors}
\end{equation}
The joint phase uses mode $\mathrm{JP}$. For unequal splitting, the subsequent
single-layer phase uses mode $\mathrm P$ when $\alpha<0.5$ and mode
$\mathrm C$ when $\alpha>0.5$.

The common-layer vector and the SCMA codewords of different users are assumed
to be zero mean and mutually independent, with
\begin{equation}
    \mathbb E\!\left[\left\|\mathbf x_{\mathrm c}^{(q)}\right\|^2\right]=K,
    \qquad
    \mathbb E\!\left[\left\|\mathbf x_j^{(q)}\right\|^2\right]=1,
    \quad j\in\mathcal J.
    \label{eq:energy_normalization}
\end{equation}
Hence, the average energy of a block transmitted in mode $\nu$ is
\begin{equation}
    \mathbb E\!\left[\left\|\mathbf x^{(q,\nu)}\right\|^2\right]
    =
    Kp_{\mathrm c}^{(\nu)}
    +
    Jp_{\mathrm p}^{(\nu)}.
    \label{eq:joint_block_energy}
\end{equation}

\subsubsection{Subcarrier Mapping and OFDM Signal Generation}

The $K$ entries of each RS-SCMA block can be mapped onto the OFDM subcarriers
using either localized or interleaved allocation. Let $\mathcal I_q$ denote
the set of subcarriers occupied by the $q$th block. Under localized mapping,
the block occupies $K$ consecutive subcarriers, i.e.,
\begin{equation}
    \mathcal I_q^{\mathrm{loc}}
    =
    \{(q-1)K+1,\ldots,qK\},
    \label{eq:localized_mapping}
\end{equation}
whereas under interleaved mapping, its entries are spaced $Q$ subcarriers
apart, i.e.,
\begin{equation}
    \mathcal I_q^{\mathrm{int}}
    =
    \{q,q+Q,\ldots,q+(K-1)Q\}.
    \label{eq:interleaved_mapping}
\end{equation}
In both cases, the $k$th entry of the block is carried on the $k$th element of
$\mathcal I_q$ in increasing order. Localized mapping confines each block to a
narrow portion of the band, over which the channel gains may be strongly
correlated, whereas interleaved mapping spreads the block across the entire
band and thus provides greater frequency diversity. Accordingly, interleaved
mapping is adopted in the remainder of this paper.

Under the adopted interleaved mapping, the $k$th entry of the $q$th RS-SCMA
block is placed on subcarrier
\begin{equation}
    n_{q,k}
    =
    (k-1)Q+q,
    \qquad
    q\in\mathcal Q,\; k\in\mathcal K.
    \label{eq:interleaved_subcarrier_index}
\end{equation}
Accordingly, the subcarrier mapping matrix of the $q$th block is defined as
\begin{equation}
    \mathbf E_q
    =
    \left[
    \mathbf e_{n_{q,1}},
    \ldots,
    \mathbf e_{n_{q,K}}
    \right]
    \in
    \{0,1\}^{N_{\mathrm{sc}}\times K},
    \label{eq:selection_matrix}
\end{equation}
where $\mathbf e_n$ denotes the $n$th column of $\mathbf I_{N_{\mathrm{sc}}}$.
Since the indices $n_{q,k}$ are distinct, $\mathbf E_q^{\mathrm H}\mathbf E_q
=\mathbf I_K$ and $\mathbf E_q^{\mathrm H}\mathbf E_{q'}=\mathbf 0$ for
$q\neq q'$. The frequency-domain transmit vector of an OFDM symbol in mode
$\nu$ is then given by
\begin{equation}
    \mathbf x_{\mathrm F}^{(\nu)}
    =
    \sum_{q=1}^{Q}
    \mathbf E_q
    \mathbf x^{(q,\nu)}
    \in
    \mathbb C^{N_{\mathrm{sc}}\times1}.
    \label{eq:ofdm_frequency_vector}
\end{equation}

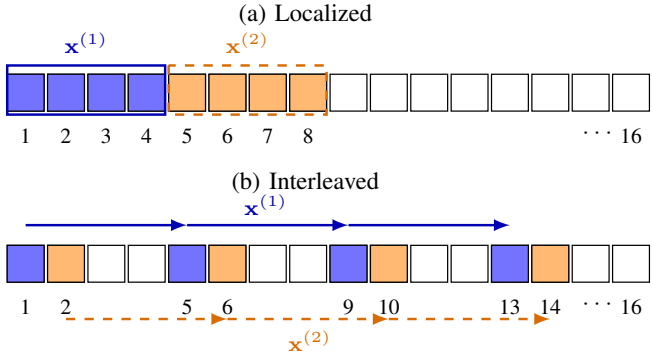
\begin{figure}[hpbt]
\centering
\resizebox{0.95\columnwidth}{!}{%
\begin{tikzpicture}[font=\scriptsize,>=latex]
\def\w{0.42}
\def\h{0.42}
\def\g{0.04}

\colorlet{blockone}{blue!55}
\colorlet{blocktwo}{orange!55}
\colorlet{oneedge}{blue!70!black}
\colorlet{twoedge}{orange!85!black}

\node[font=\footnotesize] at (3.4,1.10) {(a) Localized};

\foreach \i in {1,...,16}{
\pgfmathsetmacro{\x}{(\i-1)*(\w+\g)}
\draw[black] (\x,0) rectangle ++(\w,\h);
}

\foreach \i in {1,2,3,4}{
\pgfmathsetmacro{\x}{(\i-1)*(\w+\g)}
\fill[blockone] (\x,0) rectangle ++(\w,\h);
\draw[black] (\x,0) rectangle ++(\w,\h);
}

\foreach \i in {5,6,7,8}{
\pgfmathsetmacro{\x}{(\i-1)*(\w+\g)}
\fill[blocktwo] (\x,0) rectangle ++(\w,\h);
\draw[black] (\x,0) rectangle ++(\w,\h);
}

\draw[thick,oneedge] (0,0.52) rectangle (1.80,-0.04);
\node[text=oneedge] at (0.90,0.78) {$\mathbf{x}^{(1)}$};

\draw[thick,dashed,twoedge] (1.84,0.52) rectangle (3.64,-0.04);
\node[text=twoedge] at (2.74,0.78) {$\mathbf{x}^{(2)}$};

\foreach \i in {1,2,3,4,5,6,7,8,16}{
\pgfmathsetmacro{\x}{(\i-1)*(\w+\g)+0.5*\w}
\node[below=2pt] at (\x,0) {\i};
}
\node[below=2pt] at (6.75,0) {$\cdots$};

\node[font=\footnotesize] at (3.4,-0.82) {(b) Interleaved};

\foreach \i in {1,...,16}{
\pgfmathsetmacro{\x}{(\i-1)*(\w+\g)}
\draw[black] (\x,-1.95) rectangle ++(\w,\h);
}

\foreach \i in {1,5,9,13}{
\pgfmathsetmacro{\x}{(\i-1)*(\w+\g)}
\fill[blockone] (\x,-1.95) rectangle ++(\w,\h);
\draw[black] (\x,-1.95) rectangle ++(\w,\h);
}

\foreach \i in {2,6,10,14}{
\pgfmathsetmacro{\x}{(\i-1)*(\w+\g)}
\fill[blocktwo] (\x,-1.95) rectangle ++(\w,\h);
\draw[black] (\x,-1.95) rectangle ++(\w,\h);
}

\draw[thick,->,oneedge] (0.21,-1.30) -- (2.05,-1.30);
\draw[thick,->,oneedge] (2.05,-1.30) -- (3.89,-1.30);
\draw[thick,->,oneedge] (3.89,-1.30) -- (5.73,-1.30);
\node[text=oneedge] at (2.95,-1.07) {$\mathbf{x}^{(1)}$};

\draw[thick,dashed,->,twoedge] (0.67,-2.37) -- (2.51,-2.37);
\draw[thick,dashed,->,twoedge] (2.51,-2.37) -- (4.35,-2.37);
\draw[thick,dashed,->,twoedge] (4.35,-2.37) -- (6.19,-2.37);
\node[text=twoedge] at (3.45,-2.62) {$\mathbf{x}^{(2)}$};

\foreach \i in {1,2,5,6,9,10,13,14,16}{
\pgfmathsetmacro{\x}{(\i-1)*(\w+\g)+0.5*\w}
\node[below=2pt] at (\x,-1.95) {\i};
}
\node[below=2pt] at (6.75,-1.95) {$\cdots$};

\end{tikzpicture}%
}
\caption{Illustration of localized and interleaved subcarrier mappings for
RS-SCMA blocks with $N_{\mathrm{sc}}=16$, $K=4$, and $Q=4$. 
}
\label{fig:ofdm_mapping}
\end{figure}

For example, with $N_{\mathrm{sc}}=16$ and $K=4$, we have $Q=4$. Localized
mapping yields $\mathcal I_1^{\mathrm{loc}}=\{1,2,3,4\}$ and
$\mathcal I_2^{\mathrm{loc}}=\{5,6,7,8\}$, whereas interleaved mapping yields
$\mathcal I_1^{\mathrm{int}}=\{1,5,9,13\}$ and
$\mathcal I_2^{\mathrm{int}}=\{2,6,10,14\}$, as illustrated in
Fig.~\ref{fig:ofdm_mapping}.

The frequency-domain vector is converted to the time domain by the unitary
$N_{\mathrm{sc}}$-point inverse DFT as
\begin{equation}
    \mathbf x_{\mathrm T}^{(\nu)}
    =
    \mathbf U^{\mathrm H}
    \mathbf x_{\mathrm F}^{(\nu)}
    \in
    \mathbb C^{N_{\mathrm{sc}}\times1},
    \label{eq:ifft}
\end{equation}
where $\mathbf U\in\mathbb C^{N_{\mathrm{sc}}\times N_{\mathrm{sc}}}$ denotes
the unitary DFT matrix, whose $(n,r+1)$th entry is
\begin{equation}
    [\mathbf U]_{n,r+1}
    =
    \frac{1}{\sqrt{N_{\mathrm{sc}}}}
    e^{-i 2\pi(n-1)r/N_{\mathrm{sc}}},
    \label{eq:unitary_dft_matrix}
\end{equation}
with $n\in\mathcal N_{\mathrm{sc}}$ indexing the subcarriers and
$r\in\{0,\ldots,N_{\mathrm{sc}}-1\}$ indexing the time-domain samples.

\subsubsection{Frequency-Selective Channel and CFO}

A CP of length $N_{\mathrm{cp}}$ is appended to each OFDM
symbol. For receiving user $j$, let
\begin{equation}
    \mathbf h_j
    =
    [h_{j,0},\ldots,h_{j,L-1}]^{\mathrm T}
    \sim
    \mathcal{CN}(\mathbf 0,\mathbf R_{h,j})
    \label{eq:time_domain_channel}
\end{equation}
denote the $L$-tap discrete-time channel impulse response. The CP length
satisfies $N_{\mathrm{cp}}\geq L-1$, so that inter-symbol interference between
consecutive OFDM symbols is avoided. 

Let $\bar{\mathbf h}_j=[\mathbf h_j^{\mathrm T},
\mathbf 0_{1\times(N_{\mathrm{sc}}-L)}]^{\mathrm T}$ denote the zero-padded
channel impulse response. The corresponding frequency-domain channel vector
and diagonal channel matrix are
\begin{equation}
    \mathbf g_j
    =
    \sqrt{N_{\mathrm{sc}}}\,\mathbf U\bar{\mathbf h}_j
    =
    [g_{j,1},\ldots,g_{j,N_{\mathrm{sc}}}]^{\mathrm T},
    \label{eq:frequency_channel}
\end{equation}
\begin{equation}
    \mathbf G_j
    =
    \operatorname{diag}(\mathbf g_j),
    \label{eq:frequency_channel_matrix}
\end{equation}
respectively.  The channel is normalized such that
    $\mathbb E[|g_{j,n}|^2]=1,
    \ 
    n\in\mathcal N_{\mathrm{sc}}.$
Let $\varepsilon_j$ denote the CFO of user $j$, normalized by the subcarrier
spacing. After CP removal, its effect on the received time-domain samples is
captured by
\begin{equation}
    \mathbf D(\varepsilon_j)
    =
    \operatorname{diag}\!\left(
    1,e^{i 2\pi\varepsilon_j/N_{\mathrm{sc}}},\ldots,
    e^{i 2\pi\varepsilon_j(N_{\mathrm{sc}}-1)/N_{\mathrm{sc}}}
    \right),
    \label{eq:time_domain_cfo_matrix}
\end{equation}
The received time-domain vector at user $j$ is then
\begin{equation}
    \mathbf y_{\mathrm T,j}^{(\nu)}
    =
    \mathbf D(\varepsilon_j)
    \mathbf H_{\mathrm{circ},j}
    \mathbf x_{\mathrm T}^{(\nu)}
    +
    \mathbf w_{\mathrm T,j},
    \label{eq:received_time_domain}
\end{equation}
where $\mathbf H_{\mathrm{circ},j}$ is the circulant channel matrix generated from  $\bar{\mathbf h}_j$, and
$\mathbf w_{\mathrm T,j}\sim\mathcal{CN}(\mathbf 0,\sigma_0^2\mathbf
I_{N_{\mathrm{sc}}})$ denotes AWGN. Since
$\mathbf H_{\mathrm{circ},j}=\mathbf U^{\mathrm H}\mathbf G_j\mathbf U$,
applying the $N_{\mathrm{sc}}$-point DFT at the receiver yields
\begin{equation}
    \mathbf y_{\mathrm F,j}^{(\nu)}
    =
    \mathbf U\mathbf y_{\mathrm T,j}^{(\nu)}
    =
    \boldsymbol{\Phi}_j
    \mathbf G_j
    \mathbf x_{\mathrm F}^{(\nu)}
    +
    \mathbf w_{\mathrm F,j},
    \label{eq:user_received_ofdm_vector}
\end{equation}
where
\begin{equation}
    \boldsymbol{\Phi}_j
    =
    \mathbf U
    \mathbf D(\varepsilon_j)
    \mathbf U^{\mathrm H}
    \label{eq:user_cfo_matrix}
\end{equation}
denotes the frequency-domain CFO coupling matrix, and
$\mathbf w_{\mathrm F,j}=\mathbf U\mathbf w_{\mathrm T,j}
\sim\mathcal{CN}(\mathbf 0,\sigma_0^2\mathbf I_{N_{\mathrm{sc}}})$ since
$\mathbf U$ is unitary. For $\varepsilon_j=0$, $\boldsymbol{\Phi}_j=\mathbf
I_{N_{\mathrm{sc}}}$ and the subcarriers remain orthogonal, whereas for
$\varepsilon_j\neq0$, the off-diagonal entries of $\boldsymbol{\Phi}_j$ give
rise to ICI.

The overall signal flow from the $q$th RS-SCMA block to its received
counterpart is summarized as
\begin{equation*}
    \mathbf x^{(q,\nu)}
    \!\xrightarrow{\mathbf E_q}\!
    \mathbf x_{\mathrm F}^{(\nu)}
    \!\xrightarrow{\mathbf U^{\mathrm H}}\!
    \mathbf x_{\mathrm T}^{(\nu)}
    \!\xrightarrow{\mathcal H_j}\!
    \mathbf y_{\mathrm T,j}^{(\nu)}
    \!\xrightarrow{\mathbf U}\!
    \mathbf y_{\mathrm F,j}^{(\nu)}
    \!\xrightarrow{\mathbf E_q^{\mathrm H}}\!
    \mathbf y_j^{(q,\nu)},
\end{equation*}
where $\mathcal H_j$ represents CP insertion, propagation through the
frequency-selective channel with CFO, and CP removal.

Let $\varphi_{j;n,m}=[\boldsymbol{\Phi}_j]_{n,m}$ denote the CFO-induced
coupling coefficient from transmitted subcarrier $m$ to received subcarrier
$n$, and let $x_{\mathrm F,n}^{(\nu)}$, $y_{\mathrm F,j,n}^{(\nu)}$, and
$w_{\mathrm F,j,n}$ denote the $n$th entries of
$\mathbf x_{\mathrm F}^{(\nu)}$, $\mathbf y_{\mathrm F,j}^{(\nu)}$, and
$\mathbf w_{\mathrm F,j}$, respectively. From
\eqref{eq:user_received_ofdm_vector}, the $n$th received entry is given by
\begin{equation}
    y_{\mathrm F,j,n}^{(\nu)}
    =
    \varphi_{j;n,n}g_{j,n}x_{\mathrm F,n}^{(\nu)}
    +
    \sum_{\substack{m=1\\m\neq n}}^{N_{\mathrm{sc}}}
    \varphi_{j;n,m}g_{j,m}x_{\mathrm F,m}^{(\nu)}
    +
    w_{\mathrm F,j,n}.
    \label{eq:subcarrier_cfo_received}
\end{equation}
The first term is the desired component on subcarrier $n$, attenuated and
phase-rotated by $\varphi_{j;n,n}$, whereas the summation represents the ICI
caused by leakage from all other subcarriers through the off-diagonal
coupling coefficients.

Finally, the received components of the $q$th RS-SCMA block are extracted as
\begin{equation}
    \mathbf y_j^{(q,\nu)}
    =
    \mathbf E_q^{\mathrm H}
    \mathbf y_{\mathrm F,j}^{(\nu)}
    \in
    \mathbb C^{K\times1}.
    \label{eq:received_rsscma_block}
\end{equation}
In mode $\mathrm{JP}$, $\mathbf y_j^{(q,\mathrm{JP})}$ is processed by
common-layer detection, SIC, and MPA-based private-layer detection, whereas in
modes $\mathrm P$ and $\mathrm C$, only MPA-based private-layer detection or
common-layer detection is performed, respectively.

\subsection{Demapped RS-SCMA Block Model}
\label{subsec:demapped_block_model}

Substituting \eqref{eq:ofdm_frequency_vector} and
\eqref{eq:user_received_ofdm_vector} into \eqref{eq:received_rsscma_block},
the received $q$th RS-SCMA block at user $j$ can be expressed as
\begin{equation}
    \mathbf y_j^{(q,\nu)}
    =
    \mathbf B_j^{(q)}\mathbf x^{(q,\nu)}
    +
    \mathbf i_{\mathrm{out},j}^{(q,\nu)}
    +
    \mathbf w_j^{(q)},
    \label{eq:demapped_expanded_model}
\end{equation}
where $\mathbf w_j^{(q)}=\mathbf E_q^{\mathrm H}\mathbf w_{\mathrm F,j}
\sim\mathcal{CN}(\mathbf 0,\sigma_0^2\mathbf I_K)$,
\begin{equation}
    \mathbf B_j^{(q)}
    =
    \mathbf E_q^{\mathrm H}
    \boldsymbol{\Phi}_j
    \mathbf G_j
    \mathbf E_q
    \in
    \mathbb C^{K\times K}
    \label{eq:block_cfo_matrix}
\end{equation}
denotes the CFO-impaired effective channel of the desired block, and
\begin{equation}
    \mathbf i_{\mathrm{out},j}^{(q,\nu)}
    =
    \sum_{\substack{q'=1\\q'\neq q}}^{Q}
    \mathbf E_q^{\mathrm H}
    \boldsymbol{\Phi}_j
    \mathbf G_j
    \mathbf E_{q'}
    \mathbf x^{(q',\nu)}
    \label{eq:out_of_block_ici_system}
\end{equation}
collects the CFO-induced leakage from the remaining $Q-1$ RS-SCMA blocks of
the same OFDM symbol.

\begin{figure}[!hbpt]
\centering
\resizebox{1\columnwidth}{!}{%
\begin{tikzpicture}[
    >=Latex,
    font=\small,
    re/.style={
        draw,
        rounded corners=1pt,
        minimum width=9mm,
        minimum height=5.5mm,
        align=center
    },
    blk/.style={
        draw,
        rounded corners=2pt,
        inner sep=4pt
    },
    intra/.style={->, dashed, thick},
    outblk/.style={->, thick}
]

\node[re] (o1) at (0,0) {$\mathbf x^{(q',\nu)}$};
\node[below=2mm of o1] {\scriptsize $q'\neq q$};

\node[re] (c1) at (4.2,0) {$x_1^{(q,\nu)}$};
\node[re, right=2mm of c1] (c2) {$x_2^{(q,\nu)}$};
\node[re, right=2mm of c2] (c3) {$\cdots$};
\node[re, right=2mm of c3] (c4) {$x_K^{(q,\nu)}$};

\coordinate (ctop) at ($(c2.north)+(0,0.55)$);
\coordinate (cbot) at ($(c2.south)+(0,-0.55)$);

\node[blk, fit=(c1)(c2)(c3)(c4)(ctop)(cbot)] (blkC) {};
\node[below=3mm of blkC] {Desired block $q$};

\draw[intra]
(c1.north)
to[out=70,in=110]
(c2.north);

\draw[intra]
(c2.south)
to[out=-65,in=-120]
(c4.south);

\draw[intra]
(c4.north)
to[out=125,in=45]
(c1.north);

\node at ($(blkC.north)+(0,0.55)$)
{\footnotesize intra-block leakage};

\draw[outblk]
(o1.east)
to[out=5,in=175]
(blkC.west);

\node at ($(o1.east)!0.48!(blkC.west)+(0,0.45)$)
{\footnotesize out-of-block leakage};

\end{tikzpicture}%
}
\caption{\footnotesize CFO-induced intra-block leakage within the desired
RS-SCMA block and out-of-block leakage from the remaining blocks.}
\label{fig:cfo_intra_inter_block}
\end{figure}
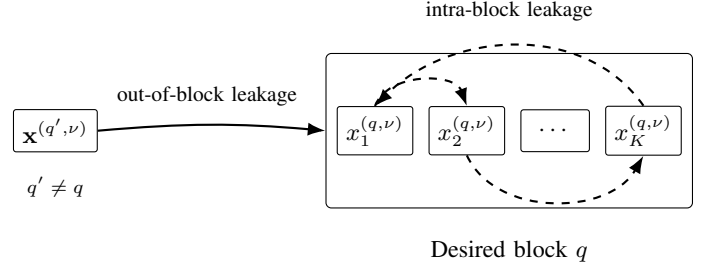

Since $\boldsymbol{\Phi}_j$ is generally nondiagonal in the presence of CFO,
$\mathbf B_j^{(q)}$ captures both the direct contribution of each RE and the
CFO-induced coupling among the $K$ REs of block $q$. Specifically, from
\eqref{eq:interleaved_subcarrier_index} and \eqref{eq:selection_matrix}, its
entries are
\begin{equation}
    [\mathbf B_j^{(q)}]_{k,k'}
    =
    \varphi_{j;n_{q,k},n_{q,k'}}\,g_{j,n_{q,k'}},
    \qquad
    k,k'\in\mathcal K.
    \label{eq:block_diagonal_offdiagonal}
\end{equation}
Accordingly, $\mathbf B_j^{(q)}$ is decomposed as
\begin{equation}
    \mathbf B_j^{(q)}
    =
    \mathbf H_j^{(q)}
    +
    \boldsymbol{\Delta}_j^{(q)},
    \label{eq:block_cfo_decomposition}
\end{equation}
where the diagonal matrix $\mathbf H_j^{(q)}$ contains the diagonal entries of
$\mathbf B_j^{(q)}$, and $\boldsymbol{\Delta}_j^{(q)}=\mathbf B_j^{(q)}-
\mathbf H_j^{(q)}$ contains the off-diagonal CFO coupling among the REs of
block $q$. Substituting \eqref{eq:block_cfo_decomposition} into
\eqref{eq:demapped_expanded_model} yields
\begin{equation}
    \mathbf y_j^{(q,\nu)}
    =
    \mathbf H_j^{(q)}
    \mathbf x^{(q,\nu)}
    +
    \boldsymbol{\Delta}_j^{(q)}
    \mathbf x^{(q,\nu)}
    +
    \mathbf i_{\mathrm{out},j}^{(q,\nu)}
    +
    \mathbf w_j^{(q)},
    \label{eq:compact_demapped_model}
\end{equation}
where the first term is the desired diagonal contribution, the second term
represents intra-block CFO leakage among the REs of block $q$, and the third
term represents out-of-block leakage from the remaining RS-SCMA blocks. These
two leakage mechanisms are illustrated in Fig.~\ref{fig:cfo_intra_inter_block}.

In the absence of CFO, i.e., $\varepsilon_j=0$, the CFO coupling matrix
reduces to $\boldsymbol{\Phi}_j=\mathbf I_{N_{\mathrm{sc}}}$, so that
$\varphi_{j;n,m}=1$ for $n=m$ and $\varphi_{j;n,m}=0$ otherwise. Since the
indices $n_{q,k}$, $k\in\mathcal K$, are distinct,
\eqref{eq:block_diagonal_offdiagonal} gives
\begin{equation}
    \mathbf B_j^{(q)}
    =
    \mathbf H_j^{(q)}
    =
    \operatorname{diag}
    \left(
    g_{j,n_{q,1}},
    \ldots,
    g_{j,n_{q,K}}
    \right),
    \label{eq:block_no_cfo_diagonal}
\end{equation}
and hence $\boldsymbol{\Delta}_j^{(q)}=\mathbf 0$. Moreover, since different
RS-SCMA blocks occupy disjoint subcarrier sets,
$\mathbf E_q^{\mathrm H}\mathbf G_j\mathbf E_{q'}=\mathbf 0$ for $q'\neq q$,
which yields $\mathbf i_{\mathrm{out},j}^{(q,\nu)}=\mathbf 0$. Consequently,
\eqref{eq:compact_demapped_model} reduces to
\begin{equation}
    \mathbf y_j^{(q,\nu)}
    =
    \mathbf H_j^{(q)}
    \mathbf x^{(q,\nu)}
    +
    \mathbf w_j^{(q)},
    \label{eq:demapped_no_cfo}
\end{equation}
i.e., the $K$ REs of each block experience parallel flat-fading subchannels,
as in conventional OFDM transmission.

\section{Error Performance Analysis of CFO-Impaired OFDM-RS-SCMA}
\label{sec:error_performance_analysis}

Based on the demapped signal model in \eqref{eq:compact_demapped_model}, this
section develops the BER analysis of the OFDM-based RS-SCMA receiver. In the
joint mode, the common layer is detected first using the diagonal effective
channel $\mathbf H_j^{(q)}$; it is then removed via SIC, after which the
private SCMA layer is detected. Since the error probability of iterative MPA
detection does not admit a closed-form expression, the private-layer BER is
analyzed through the pairwise error probabilities (PEPs) of the corresponding
ML detection metric. In this analysis, the desired
finite-alphabet common and private components conveyed through
$\mathbf H_j^{(q)}$ are retained explicitly, whereas the unresolved
CFO-induced leakage, comprising the intra-block term
$\boldsymbol{\Delta}_j^{(q)}\mathbf x^{(q,\nu)}$ and the out-of-block term
$\mathbf i_{\mathrm{out},j}^{(q,\nu)}$, is approximated as a correlated
Gaussian disturbance characterized by its second-order statistics conditioned
on the channel realization. Together with the AWGN, this leakage constitutes
the effective disturbance used in the subsequent PEP analysis.

\subsection{Desired-Block Signal and Interference Model}
\label{subsec:desired_block_model_general_alpha}

For the error analysis, consider an arbitrary receiving user $j$ and RS-SCMA
block $q$. The error events can be evaluated jointly over the $K$ REs of this
block. For notational simplicity, the user and block indices are dropped,
i.e., $\mathbf B=\mathbf B_j^{(q)}$, $\mathbf H=\mathbf H_j^{(q)}$, and
$\boldsymbol{\Delta}=\boldsymbol{\Delta}_j^{(q)}$, and similarly for
$\mathbf i_{\mathrm{out}}^{(\nu)}$ and $\mathbf w$.

Let $\mathcal A_{\mathrm c}=\{a_1,\ldots,a_M\}$. The transmitted common-layer
vector is indexed by $\mathbf u=(u_1,\ldots,u_K)\in\{1,\ldots,M\}^K$ as
\begin{equation}
    \mathbf x_{\mathrm c,\mathbf u}
    =
    \left[
    a_{u_1},
    \ldots,
    a_{u_K}
    \right]^{\mathrm T}
    \in
    \mathcal A_{\mathrm c}^{K}.
    \label{eq:indexed_common_vector}
\end{equation}
Similarly, the private layer is indexed by the codeword-index vector
$\mathbf v=(v_1,\ldots,v_J)\in\{1,\ldots,M\}^J$, and the aggregate
private-layer vector is
\begin{equation}
    \mathbf x_{\mathrm p,\mathbf v}
    =
    \sum_{j'=1}^{J}
    \mathbf c_{j',v_{j'}},
    \label{eq:indexed_private_vector}
\end{equation}
where $\mathbf c_{j',v_{j'}}\in\mathcal C_{j'}$ is the $v_{j'}$th codeword of user $j'$.

In the joint mode, i.e., $\nu=\mathrm{JP}$, the demapped received block is
\begin{equation}
\begin{aligned}
\mathbf y^{(\mathrm{JP})}
={}&
(\mathbf H+\boldsymbol{\Delta})
\left(
\sqrt{p_{\mathrm c}^{(\mathrm{JP})}}\,
\mathbf x_{\mathrm c,\mathbf u}
+
\sqrt{p_{\mathrm p}^{(\mathrm{JP})}}\,
\mathbf x_{\mathrm p,\mathbf v}
\right)
\\
&+
\mathbf i_{\mathrm{out}}^{(\mathrm{JP})}
+
\mathbf w,
\end{aligned}
\label{eq:analysis_received_joint_block}
\end{equation}
where $\mathbf w\sim\mathcal{CN}(\mathbf 0,\sigma_0^2\mathbf I_K)$. The
common and private-layer components conveyed through $\mathbf H$ are
retained explicitly, whereas the components coupled through
$\boldsymbol{\Delta}$, the out-of-block leakage, and the AWGN are collected
into the effective disturbance
\begin{equation}
\begin{aligned}
\mathbf z_{\mathbf u,\mathbf v}^{(\mathrm{JP})}
={}&
\sqrt{p_{\mathrm c}^{(\mathrm{JP})}}\,
\boldsymbol{\Delta}
\mathbf x_{\mathrm c,\mathbf u}
+
\sqrt{p_{\mathrm p}^{(\mathrm{JP})}}\,
\boldsymbol{\Delta}
\mathbf x_{\mathrm p,\mathbf v}
\\
&+
\mathbf i_{\mathrm{out}}^{(\mathrm{JP})}
+
\mathbf w.
\end{aligned}
\label{eq:effective_disturbance_joint}
\end{equation}
The model used for the joint-mode error analysis is therefore
\begin{equation}
    \mathbf y^{(\mathrm{JP})}
    =
    \sqrt{p_{\mathrm c}^{(\mathrm{JP})}}\,
    \mathbf H
    \mathbf x_{\mathrm c,\mathbf u}
    +
    \sqrt{p_{\mathrm p}^{(\mathrm{JP})}}\,
    \mathbf H
    \mathbf x_{\mathrm p,\mathbf v}
    +
    \mathbf z_{\mathbf u,\mathbf v}^{(\mathrm{JP})}.
    \label{eq:joint_analysis_model}
\end{equation}

\subsection{Correlated CFO-Induced Interference Covariance}
\label{subsec:correlated_interference_covariance}

The common-layer symbols are assumed to be independent and zero mean with unit
average energy. For the private layer, each SCMA codeword has unit average
energy, so the aggregate private-layer vector has average energy $J$. Assuming
that this energy is uniformly distributed across the $K$ REs and neglecting the
correlation between the nonzero entries of each codeword, we have
\begin{equation}
    \mathbb E\!\left[
    \mathbf x_{\mathrm c,\mathbf u}
    \mathbf x_{\mathrm c,\mathbf u}^{\mathrm H}
    \right]
    =
    \mathbf I_K,
    \qquad
    \mathbb E\!\left[
    \mathbf x_{\mathrm p,\mathbf v}
    \mathbf x_{\mathrm p,\mathbf v}^{\mathrm H}
    \right]
    \approx
    \frac{J}{K}\mathbf I_K.
    \label{eq:layer_covariances}
\end{equation}
The second approximation is used only to characterize the unresolved
CFO-induced leakage. From \eqref{eq:interval_transmit_vectors} and the
independence of the two layers, the transmit covariance of a block in mode
$\nu\in\{\mathrm{JP},\mathrm P,\mathrm C\}$ is then
\begin{equation}
    \mathbb E\!\left[
    \mathbf x^{(\nu)}
    \left(\mathbf x^{(\nu)}\right)^{\mathrm H}
    \right]
    \approx
    E_{\mathrm{RE}}^{(\nu)}
    \mathbf I_K,
    \label{eq:desired_block_symbol_covariance}
\end{equation}
where
\begin{equation}
    E_{\mathrm{RE}}^{(\nu)}
    =
    p_{\mathrm c}^{(\nu)}
    +
    \frac{J}{K}p_{\mathrm p}^{(\nu)}
    \label{eq:interval_per_re_energy}
\end{equation}
denotes the average energy per RE, with
$p_{\mathrm c}^{(\mathrm P)}=p_{\mathrm p}^{(\mathrm C)}=0$.

To characterize the out-of-block leakage, the block index $q$ is restored in
the subcarrier mapping matrices. Let
\begin{equation}
    \mathbf E_{\mathrm{out}}
    =
    \left[
    \mathbf E_1,\ldots,\mathbf E_{q-1},
    \mathbf E_{q+1},\ldots,\mathbf E_Q
    \right]
    \label{eq:Eout_definition}
\end{equation}
collect the mapping matrices of the remaining $Q-1$ blocks, and let
$\mathbf x_{\mathrm{out}}^{(\nu)}\in\mathbb C^{(Q-1)K\times1}$ stack their
transmitted vectors $\mathbf x^{(q',\nu)}$, $q'\neq q$, in the same order.
The out-of-block leakage can then be written as
\begin{equation}
    \mathbf i_{\mathrm{out}}^{(\nu)}
    =
    \mathbf A_{\mathrm{out}}(\mathbf g)
    \mathbf x_{\mathrm{out}}^{(\nu)},
    \label{eq:outblock_vector_matrix_form}
\end{equation}
where
\begin{equation}
    \mathbf A_{\mathrm{out}}(\mathbf g)
    =
    \mathbf E_q^{\mathrm H}
    \boldsymbol{\Phi}
    \mathbf G
    \mathbf E_{\mathrm{out}}
    \in
    \mathbb C^{K\times (Q-1)K}
    \label{eq:Aout_definition}
\end{equation}
describes the CFO coupling from the remaining blocks to block $q$. Since all
blocks are generated independently with the same normalization,
$\mathbb E[\mathbf x_{\mathrm{out}}^{(\nu)}
(\mathbf x_{\mathrm{out}}^{(\nu)})^{\mathrm H}]
\approx E_{\mathrm{RE}}^{(\nu)}\mathbf I_{(Q-1)K}$. Hence, conditioned on the
channel realization $\mathbf g$, the covariance of the out-of-block leakage is
\begin{equation}
    \mathbf R_I^{(\nu)}(\mathbf g)
    =
    \mathbb E\!\left[
    \mathbf i_{\mathrm{out}}^{(\nu)}
    \left(\mathbf i_{\mathrm{out}}^{(\nu)}\right)^{\mathrm H}
    \,\middle|\,
    \mathbf g
    \right]
    \approx
    E_{\mathrm{RE}}^{(\nu)}
    \mathbf A_{\mathrm{out}}(\mathbf g)
    \mathbf A_{\mathrm{out}}^{\mathrm H}(\mathbf g),
    \label{eq:outblock_covariance}
\end{equation}
which retains the correlation induced by CFO leakage across the $K$ REs of
block $q$.

Similarly, the intra-block leakage is
$\mathbf i_{\mathrm{in}}^{(\nu)}=\boldsymbol{\Delta}\mathbf x^{(\nu)}$.
Averaging over the transmitted symbols of block $q$ using
\eqref{eq:desired_block_symbol_covariance}, its conditional covariance is
\begin{equation}
    \mathbf R_{\Delta}^{(\nu)}(\mathbf g)
    =
    \mathbb E\!\left[
    \mathbf i_{\mathrm{in}}^{(\nu)}
    \left(\mathbf i_{\mathrm{in}}^{(\nu)}\right)^{\mathrm H}
    \,\middle|\,
    \mathbf g
    \right]
    \approx
    E_{\mathrm{RE}}^{(\nu)}
    \boldsymbol{\Delta}
    \boldsymbol{\Delta}^{\mathrm H}.
    \label{eq:inblock_covariance}
\end{equation}
Since the transmitted vectors of different blocks are independent,
$\mathbf i_{\mathrm{in}}^{(\nu)}$ and $\mathbf i_{\mathrm{out}}^{(\nu)}$ are
uncorrelated. Including the AWGN, the conditional covariance of the effective
disturbance is therefore
\begin{equation}
    \mathbf R_{\mathrm{eff}}^{(\nu)}(\mathbf g)
    =
    \mathbf R_I^{(\nu)}(\mathbf g)
    +
    \mathbf R_{\Delta}^{(\nu)}(\mathbf g)
    +
    \sigma_0^2\mathbf I_K.
    \label{eq:R0_interval_def}
\end{equation}

Since the CFO-induced leakage depends on finite-alphabet symbols, the
effective disturbance is not Gaussian in general. For tractable PEP analysis,
it is modeled as a proper complex Gaussian vector with the same conditional
covariance, i.e.,
$\mathbf z^{(\nu)}\sim\mathcal{CN}(\mathbf 0,
\mathbf R_{\mathrm{eff}}^{(\nu)}(\mathbf g))$, independent of the desired
finite-alphabet components of block $q$.

\subsection{Joint-Phase Common-Layer PEP}
\label{subsec:common_layer_pep_general_alpha}

In the joint mode, i.e., $\nu=\mathrm{JP}$, the common layer is detected first
while the private SCMA layer remains superimposed. Using the diagonal
effective channel $\mathbf H$, the common-layer decision is
\begin{equation}
    \widehat{\mathbf u}
    =
    \arg\min_{\widetilde{\mathbf u}\in\{1,\ldots,M\}^K}
    \left\|
    \mathbf y^{(\mathrm{JP})}
    -
    \sqrt{p_{\mathrm c}^{(\mathrm{JP})}}\,
    \mathbf H
    \mathbf x_{\mathrm c,\widetilde{\mathbf u}}
    \right\|^2 .
    \label{eq:common_receiver_metric_joint}
\end{equation}
Since the private layer has not yet been cancelled, its finite-alphabet
contribution is retained explicitly as
\begin{equation}
    \mathbf r_{\mathrm c}^{(\mathrm{JP})}(\mathbf v)
    =
    \sqrt{p_{\mathrm p}^{(\mathrm{JP})}}\,
    \mathbf H
    \mathbf x_{\mathrm p,\mathbf v},
    \label{eq:common_private_interference}
\end{equation}
so that \eqref{eq:joint_analysis_model} becomes
\begin{equation}
    \mathbf y^{(\mathrm{JP})}
    =
    \sqrt{p_{\mathrm c}^{(\mathrm{JP})}}\,
    \mathbf H
    \mathbf x_{\mathrm c,\mathbf u}
    +
    \mathbf r_{\mathrm c}^{(\mathrm{JP})}(\mathbf v)
    +
    \mathbf z^{(\mathrm{JP})},
    \label{eq:common_effective_observation}
\end{equation}
where $\mathbf z^{(\mathrm{JP})}\sim\mathcal{CN}(\mathbf 0,
\mathbf R_{\mathrm{eff}}^{(\mathrm{JP})}(\mathbf g))$ as in
Section~\ref{subsec:correlated_interference_covariance}.

Consider the pairwise error event $\mathbf u\rightarrow\mathbf u'$, with
$\mathbf u'\neq\mathbf u$, and define the difference vector
\begin{equation}
    \mathbf d_{\mathrm c}^{(\mathrm{JP})}
    =
    \sqrt{p_{\mathrm c}^{(\mathrm{JP})}}\,
    \mathbf H
    \left(
    \mathbf x_{\mathrm c,\mathbf u}
    -
    \mathbf x_{\mathrm c,\mathbf u'}
    \right),
    \label{eq:common_pairwise_difference}
\end{equation}
whose dependence on $(\mathbf u,\mathbf u')$ is suppressed for brevity.
Substituting \eqref{eq:common_effective_observation} into
\eqref{eq:common_receiver_metric_joint}, this event occurs if and only if
\begin{equation}
\begin{aligned}
&
\left\|
\mathbf d_{\mathrm c}^{(\mathrm{JP})}
\right\|^2
+
2\Re\!\left\{
\left(
\mathbf d_{\mathrm c}^{(\mathrm{JP})}
\right)^{\mathrm H}
\mathbf r_{\mathrm c}^{(\mathrm{JP})}(\mathbf v)
\right\}
\\
&\quad+
2\Re\!\left\{
\left(
\mathbf d_{\mathrm c}^{(\mathrm{JP})}
\right)^{\mathrm H}
\mathbf z^{(\mathrm{JP})}
\right\}
\leq 0 .
\end{aligned}
\label{eq:common_pairwise_scalar_event}
\end{equation}
Conditioned on $\mathbf g$ and $\mathbf v$, the last term on the left-hand
side of \eqref{eq:common_pairwise_scalar_event} is a zero-mean real Gaussian
random variable with variance
\begin{equation}
    \left(\sigma_{\mathrm c}^{(\mathrm{JP})}\right)^2
    =
    2
    \left(
    \mathbf d_{\mathrm c}^{(\mathrm{JP})}
    \right)^{\mathrm H}
    \mathbf R_{\mathrm{eff}}^{(\mathrm{JP})}(\mathbf g)
    \mathbf d_{\mathrm c}^{(\mathrm{JP})},
    \label{eq:common_pep_variance}
\end{equation}
whereas the first two terms are deterministic and sum to
\begin{equation}
    \mu_{\mathrm c}^{(\mathrm{JP})}(\mathbf v)
    =
    \left\|
    \mathbf d_{\mathrm c}^{(\mathrm{JP})}
    \right\|^2
    +
    2\Re\!\left\{
    \left(
    \mathbf d_{\mathrm c}^{(\mathrm{JP})}
    \right)^{\mathrm H}
    \mathbf r_{\mathrm c}^{(\mathrm{JP})}(\mathbf v)
    \right\}.
    \label{eq:common_pep_mean}
\end{equation}
Hence, the conditional common-layer PEP is approximated as
\begin{equation}
    P_{\mathrm c}^{(\mathrm{JP})}
    \left(
    \mathbf u\rightarrow\mathbf u'
    \mid
    \mathbf g,\mathbf v
    \right)
    \approx
    \mathrm Q\!\left(
    \frac{\mu_{\mathrm c}^{(\mathrm{JP})}(\mathbf v)}
    {\sigma_{\mathrm c}^{(\mathrm{JP})}}
    \right),
    \label{eq:common_conditional_pep_joint}
\end{equation}
where $\mathrm Q(\cdot)$ denotes the Gaussian Q-function.
\subsection{Joint-Phase SIC and Private-Layer PEP}
\label{subsec:private_layer_pep_general_alpha}

After common-layer detection, the detected common signal is reconstructed and
removed through SIC. For transmitted and detected common-symbol index vectors
$\mathbf u$ and $\widehat{\mathbf u}$, respectively, the SIC residual is
\begin{equation}
    \mathbf r_{\mathrm{sic}}^{(\mathrm{JP})}
    (\mathbf u,\widehat{\mathbf u})
    =
    \sqrt{p_{\mathrm c}^{(\mathrm{JP})}}\,
    \mathbf H
    \left(
    \mathbf x_{\mathrm c,\mathbf u}
    -
    \mathbf x_{\mathrm c,\widehat{\mathbf u}}
    \right),
    \label{eq:sic_residual_joint}
\end{equation}
which vanishes if $\widehat{\mathbf u}=\mathbf u$. Subtracting
$\sqrt{p_{\mathrm c}^{(\mathrm{JP})}}\,\mathbf H\mathbf x_{\mathrm c,
\widehat{\mathbf u}}$ from \eqref{eq:common_effective_observation} yields the
post-SIC observation
\begin{equation}
    \widetilde{\mathbf y}^{(\mathrm{JP})}
    =
    \sqrt{p_{\mathrm p}^{(\mathrm{JP})}}\,
    \mathbf H\mathbf x_{\mathrm p,\mathbf v}
    +
    \mathbf r_{\mathrm{sic}}^{(\mathrm{JP})}
    (\mathbf u,\widehat{\mathbf u})
    +
    \mathbf z^{(\mathrm{JP})}.
    \label{eq:post_sic_effective_observation}
\end{equation}
The private layer is then detected according to the metric
\begin{equation}
    \widehat{\mathbf v}
    =
    \arg\min_{\widetilde{\mathbf v}\in\{1,\ldots,M\}^J}
    \left\|
    \widetilde{\mathbf y}^{(\mathrm{JP})}
    -
    \sqrt{p_{\mathrm p}^{(\mathrm{JP})}}\,
    \mathbf H
    \mathbf x_{\mathrm p,\widetilde{\mathbf v}}
    \right\|^2,
    \label{eq:private_receiver_metric_joint}
\end{equation}
which the MPA approximates (equally-likely input) with reduced complexity by exploiting the sparse
factor graph.

Consider the pairwise error event $\mathbf v\rightarrow\mathbf v'$, with
$\mathbf v'\neq\mathbf v$, and define the difference vector
\begin{equation}
    \mathbf d_{\mathrm p}^{(\mathrm{JP})}
    =
    \sqrt{p_{\mathrm p}^{(\mathrm{JP})}}\,
    \mathbf H
    \left(
    \mathbf x_{\mathrm p,\mathbf v}
    -
    \mathbf x_{\mathrm p,\mathbf v'}
    \right).
    \label{eq:private_pairwise_difference_joint}
\end{equation}
Substituting \eqref{eq:post_sic_effective_observation} into
\eqref{eq:private_receiver_metric_joint}, this event occurs if and only if
\begin{equation}
\begin{aligned}
&
\left\|
\mathbf d_{\mathrm p}^{(\mathrm{JP})}
\right\|^2
+
2\Re\!\left\{
\left(
\mathbf d_{\mathrm p}^{(\mathrm{JP})}
\right)^{\mathrm H}
\mathbf r_{\mathrm{sic}}^{(\mathrm{JP})}
(\mathbf u,\widehat{\mathbf u})
\right\}
\\
&\quad+
2\Re\!\left\{
\left(
\mathbf d_{\mathrm p}^{(\mathrm{JP})}
\right)^{\mathrm H}
\mathbf z^{(\mathrm{JP})}
\right\}
\leq 0 .
\end{aligned}
\label{eq:private_pairwise_scalar_event_joint}
\end{equation}
Proceeding as in \eqref{eq:common_pep_variance}--%
\eqref{eq:common_conditional_pep_joint}, and neglecting the statistical
dependence between $\widehat{\mathbf u}$ and $\mathbf z^{(\mathrm{JP})}$, the
conditional private-layer PEP is approximated as
\begin{equation}
    P_{\mathrm p}^{(\mathrm{JP})}
    \left(
    \mathbf v\rightarrow\mathbf v'
    \mid
    \mathbf g,\mathbf u,\widehat{\mathbf u}
    \right)
    \approx
    \mathrm Q\!\left(
    \frac{\mu_{\mathrm p}^{(\mathrm{JP})}(\mathbf u,\widehat{\mathbf u})}
    {\sigma_{\mathrm p}^{(\mathrm{JP})}}
    \right),
    \label{eq:private_conditional_pep_joint}
\end{equation}
where
\begin{equation}
    \mu_{\mathrm p}^{(\mathrm{JP})}(\mathbf u,\widehat{\mathbf u})
    =
    \left\|
    \mathbf d_{\mathrm p}^{(\mathrm{JP})}
    \right\|^2
    +
    2\Re\!\left\{
    \left(
    \mathbf d_{\mathrm p}^{(\mathrm{JP})}
    \right)^{\mathrm H}
    \mathbf r_{\mathrm{sic}}^{(\mathrm{JP})}
    (\mathbf u,\widehat{\mathbf u})
    \right\}
    \label{eq:private_pep_mean}
\end{equation}
and
$\big(\sigma_{\mathrm p}^{(\mathrm{JP})}\big)^2
=2\big(\mathbf d_{\mathrm p}^{(\mathrm{JP})}\big)^{\mathrm H}
\mathbf R_{\mathrm{eff}}^{(\mathrm{JP})}(\mathbf g)
\mathbf d_{\mathrm p}^{(\mathrm{JP})}$.

To account for common-to-private SIC error propagation, the private-layer PEP
is averaged over the common-layer decisions as
\begin{equation}
\begin{aligned}
    & P_{\mathrm p}^{(\mathrm{JP})}
    \left(
    \mathbf v\rightarrow\mathbf v'
    \mid\mathbf g,\mathbf u
    \right)  \\
    & \qquad =
    \sum_{\widehat{\mathbf u}}
    P\!\left(
    \widehat{\mathbf u}\mid\mathbf u,\mathbf v,\mathbf g
    \right)
    P_{\mathrm p}^{(\mathrm{JP})}
    \left(
    \mathbf v\rightarrow\mathbf v'
    \mid
    \mathbf g,\mathbf u,\widehat{\mathbf u}
    \right).
    \end{aligned}
    \label{eq:private_pep_common_average}
\end{equation}
Since $\mathbf H$ is diagonal, \eqref{eq:common_receiver_metric_joint}
decouples into $K$ per-RE decisions. Assuming that these decisions are
conditionally independent given $\mathbf g$ and $\mathbf v$, and retaining
only the all-correct and single-error common-layer events,
\eqref{eq:private_pep_common_average} is approximated as
\begin{equation}
\begin{aligned}
&P_{\mathrm p}^{(\mathrm{JP})}
\left(
\mathbf v\rightarrow\mathbf v'
\mid\mathbf g,\mathbf u
\right)
\approx
\prod_{k=1}^{K}
\left(
1-P_{\mathrm e,\mathrm c,k}^{(\mathrm{JP})}
\right)
\mathrm Q\!\left(
\frac{\mu_{\mathrm p}^{(\mathrm{JP})}(\mathbf u,\mathbf u)}
{\sigma_{\mathrm p}^{(\mathrm{JP})}}
\right)
\\
&\quad+
\sum_{k=1}^{K}
\prod_{\substack{k'=1\\k'\neq k}}^{K}
\left(
1-P_{\mathrm e,\mathrm c,k'}^{(\mathrm{JP})}
\right)\\
& \quad \times  
\sum_{\widehat u_k\neq u_k}
P\!\left(
\widehat u_k\mid u_k,\mathbf v,\mathbf g
\right)
\mathrm Q\!\left(
\frac{\mu_{\mathrm p}^{(\mathrm{JP})}(\mathbf u,\widehat{\mathbf u}^{(k)})}
{\sigma_{\mathrm p}^{(\mathrm{JP})}}
\right),
\end{aligned}
\label{eq:private_pep_sic_mixture}
\end{equation}
where
$P_{\mathrm e,\mathrm c,k}^{(\mathrm{JP})}=\sum_{\widehat u_k\neq u_k}
P(\widehat u_k\mid u_k,\mathbf v,\mathbf g)$ is the common-layer symbol error
probability on RE $k$, and $\widehat{\mathbf u}^{(k)}$ denotes the index vector
obtained from $\mathbf u$ by replacing $u_k$ with $\widehat u_k$. In the first
term, $\mu_{\mathrm p}^{(\mathrm{JP})}(\mathbf u,\mathbf u)=
\|\mathbf d_{\mathrm p}^{(\mathrm{JP})}\|^2$, since the SIC residual vanishes.
In the second term, the residual in \eqref{eq:sic_residual_joint} has a single
nonzero entry,
$\sqrt{p_{\mathrm c}^{(\mathrm{JP})}}\,[\mathbf H]_{k,k}
(a_{u_k}-a_{\widehat u_k})$, at position $k$. This truncation is accurate
when the per-RE common-layer error probabilities are small.

\subsection{Single-Layer-Interval PEPs}
\label{subsec:single_layer_intervals_general_alpha}

For unequal message splitting, the joint phase is followed by a single-layer
phase in which only the private layer ($\nu=\mathrm P$, $\alpha<0.5$) or only
the common layer ($\nu=\mathrm C$, $\alpha>0.5$) is active. Consequently,
neither cross-layer interference nor SIC error propagation is present, and the
observation models reduce to
\begin{equation}
    \mathbf y^{(\mathrm P)}
    =
    \sqrt{p_{\mathrm p}^{(\mathrm P)}}\,
    \mathbf H
    \mathbf x_{\mathrm p,\mathbf v}
    +
    \mathbf z^{(\mathrm P)},
    \label{eq:private_only_observation}
\end{equation}
\begin{equation}
    \mathbf y^{(\mathrm C)}
    =
    \sqrt{p_{\mathrm c}^{(\mathrm C)}}\,
    \mathbf H
    \mathbf x_{\mathrm c,\mathbf u}
    +
    \mathbf z^{(\mathrm C)},
    \label{eq:common_only_observation}
\end{equation}
where $\mathbf z^{(\nu)}\sim\mathcal{CN}(\mathbf 0,
\mathbf R_{\mathrm{eff}}^{(\nu)}(\mathbf g))$. Following the steps in
\eqref{eq:common_pairwise_difference}--\eqref{eq:common_conditional_pep_joint}
with the cross-layer and SIC residual terms set to zero, the conditional PEPs
are
\begin{equation}
    P_{\mathrm p}^{(\mathrm P)}
    \left(
    \mathbf v\rightarrow\mathbf v'
    \mid\mathbf g
    \right)
    \approx
    \mathrm Q\!\left(
    \frac{\|\mathbf d_{\mathrm p}^{(\mathrm P)}\|^2}
    {\sigma_{\mathrm p}^{(\mathrm P)}}
    \right),
    \label{eq:private_only_conditional_pep}
\end{equation}
\begin{equation}
    P_{\mathrm c}^{(\mathrm C)}
    \left(
    \mathbf u\rightarrow\mathbf u'
    \mid\mathbf g
    \right)
    \approx
    \mathrm Q\!\left(
    \frac{\|\mathbf d_{\mathrm c}^{(\mathrm C)}\|^2}
    {\sigma_{\mathrm c}^{(\mathrm C)}}
    \right),
    \label{eq:common_only_conditional_pep}
\end{equation}
where $\mathbf d_{\mathrm p}^{(\mathrm P)}=\sqrt{p_{\mathrm p}^{(\mathrm P)}}
\,\mathbf H(\mathbf x_{\mathrm p,\mathbf v}-\mathbf x_{\mathrm p,\mathbf v'})$
and $\mathbf d_{\mathrm c}^{(\mathrm C)}=\sqrt{p_{\mathrm c}^{(\mathrm C)}}
\,\mathbf H(\mathbf x_{\mathrm c,\mathbf u}-\mathbf x_{\mathrm c,\mathbf u'})$
are the corresponding difference vectors, and
$\big(\sigma_{\mathrm p}^{(\mathrm P)}\big)^2$ and
$\big(\sigma_{\mathrm c}^{(\mathrm C)}\big)^2$ are defined as in
\eqref{eq:common_pep_variance} with $\mathbf R_{\mathrm{eff}}^{(\mathrm P)}
(\mathbf g)$ and $\mathbf R_{\mathrm{eff}}^{(\mathrm C)}(\mathbf g)$,
respectively. Although the cross-layer interference and SIC residual are
absent, CFO-induced leakage persists through
$\mathbf R_{\mathrm{eff}}^{(\nu)}(\mathbf g)$, whose form is unchanged except
that the leakage level scales with $E_{\mathrm{RE}}^{(\nu)}$.

In the special case $\alpha=0$, only the private layer is transmitted, and the
system reduces to the OFDM-SCMA system with CFO considered in
\cite{ofdm_scma}; its private-layer PEP is then given by
\eqref{eq:private_only_conditional_pep}.

\subsection{Fading-Averaged Layer-Wise and Overall BER Analysis}
\label{subsec:fading_averaged_ber_general_alpha}

Each common-layer vector carries $B_{\mathrm c}=K\log_2 M$ bits, whereas each
aggregate private-layer vector carries $B_{\mathrm p}=J\log_2 M$ bits. Let
$\boldsymbol{\beta}_{\mathrm c}(\mathbf u)$ and
$\boldsymbol{\beta}_{\mathrm p}(\mathbf v)$ denote the bit labels of
$\mathbf u$ and $\mathbf v$, respectively, and let $d_{\mathrm H}(\cdot,\cdot)$
denote the Hamming distance. All common-layer symbols and SCMA codewords are
assumed to be equiprobable.

For tractability, the union of pairwise error events is restricted to nearest
competitors. For the common layer, $\mathcal N_{\mathrm c}(\mathbf u)$ contains
the index vectors that differ from $\mathbf u$ in a single RE $k$, in which
$a_{u_k}$ is replaced by one of its nearest constellation neighbors. With Gray
mapping, each such competitor differs from $\mathbf u$ in one bit, i.e.,
$d_{\mathrm H}=1$. For the private layer, $\mathcal N_{\mathrm p}(\mathbf v)$
contains the $N_{\mathrm{nn}}$ aggregate vectors $\mathbf x_{\mathrm p,\mathbf
v'}$, $\mathbf v'\neq\mathbf v$, closest to $\mathbf x_{\mathrm p,\mathbf v}$
in Euclidean distance among the $M^J-1$ alternatives; their Hamming distances
follow from the users' bit labels and may exceed one. In the numerical
evaluation, $N_{\mathrm{nn}}=64$ is used, since larger values were found to
change the evaluated BER negligibly. This truncation is applied only to the
analytical BER; the Monte Carlo simulations employ MPA detection without
truncation.

Averaging the conditional PEPs over the channel and the transmitted vectors,
the joint-phase BERs of the common and private layers are approximated as
\begin{equation}
\begin{aligned}
\mathrm{BER}_{\mathrm c}^{(\mathrm{JP})}
\approx{}&
\frac{1}{B_{\mathrm c}}
\mathbb E_{\mathbf g,\mathbf u,\mathbf v}
\Bigg[
\sum_{\mathbf u'\in\mathcal N_{\mathrm c}(\mathbf u)}
d_{\mathrm H}
\left(
\boldsymbol{\beta}_{\mathrm c}(\mathbf u),
\boldsymbol{\beta}_{\mathrm c}(\mathbf u')
\right)
\\[-1mm]
&\qquad\qquad\times
P_{\mathrm c}^{(\mathrm{JP})}
\left(
\mathbf u\rightarrow\mathbf u'
\mid
\mathbf g,\mathbf v
\right)
\Bigg],
\end{aligned}
\label{eq:common_ber_joint_general_alpha}
\end{equation}
\begin{equation}
\begin{aligned}
\mathrm{BER}_{\mathrm p}^{(\mathrm{JP})}
\approx{}&
\frac{1}{B_{\mathrm p}}
\mathbb E_{\mathbf g,\mathbf u,\mathbf v}
\Bigg[
\sum_{\mathbf v'\in\mathcal N_{\mathrm p}(\mathbf v)}
d_{\mathrm H}
\left(
\boldsymbol{\beta}_{\mathrm p}(\mathbf v),
\boldsymbol{\beta}_{\mathrm p}(\mathbf v')
\right)
\\[-1mm]
&\qquad\qquad\times
P_{\mathrm p}^{(\mathrm{JP})}
\left(
\mathbf v\rightarrow\mathbf v'
\mid
\mathbf g,\mathbf u
\right)
\Bigg],
\end{aligned}
\label{eq:private_ber_joint_general_alpha}
\end{equation}
where the private-layer PEP is the SIC-averaged PEP in
\eqref{eq:private_pep_sic_mixture}. The single-layer BERs
$\mathrm{BER}_{\mathrm p}^{(\mathrm P)}$ and
$\mathrm{BER}_{\mathrm c}^{(\mathrm C)}$ are obtained from
\eqref{eq:private_ber_joint_general_alpha} and
\eqref{eq:common_ber_joint_general_alpha} by replacing the conditional PEPs
with \eqref{eq:private_only_conditional_pep} and
\eqref{eq:common_only_conditional_pep}, respectively, and removing the
expectation over the index vector of the absent layer.

The overall BER is obtained by weighting the BER of each layer and phase by
the number of bits it conveys. From \eqref{min_l} and \eqref{min_alpha}, the
durations of the joint and single-layer phases normalized by $N$ are $\gamma_{\mathrm{JP}}
=
\min\{\alpha,1-\alpha\},
\ 
\gamma_{\mathrm S}
=
|1-2\alpha|$ respectively.
 The corresponding total bit weight is
$B_{\mathrm{tot}}
=
\gamma_{\mathrm{JP}}
(B_{\mathrm c}+B_{\mathrm p})
+
\gamma_{\mathrm S}B_{\mathrm{dom}}.$
Then,
\begin{equation}
\begin{aligned}
    & \mathrm{BER}_{\mathrm o}(\alpha)
    = \\
    & \quad 
    \frac{\gamma_{\mathrm{JP}}B_{\mathrm c}}{B_{\mathrm{tot}}}
    \mathrm{BER}_{\mathrm c}^{(\mathrm{JP})}
    +
    \frac{\gamma_{\mathrm{JP}}B_{\mathrm p}}{B_{\mathrm{tot}}}
    \mathrm{BER}_{\mathrm p}^{(\mathrm{JP})}
    +
    \frac{\gamma_{\mathrm S}B_{\mathrm{dom}}}{B_{\mathrm{tot}}}
    \mathrm{BER}_{\mathrm{dom}},
    \end{aligned}
    \label{eq:overall_ber}
\end{equation}
where $(B_{\mathrm{dom}},\mathrm{BER}_{\mathrm{dom}})=(B_{\mathrm p},
\mathrm{BER}_{\mathrm p}^{(\mathrm P)})$ for $\alpha<0.5$ and
$(B_{\mathrm c},\mathrm{BER}_{\mathrm c}^{(\mathrm C)})$ for $\alpha>0.5$. For
$\alpha=0.5$, $\gamma_{\mathrm S}=0$, and \eqref{eq:overall_ber} reduces to the
bit-weighted average of the two joint-phase BERs.

\section{Simulation Results}
\label{sec:results}

This section compares the analytical BER results with Monte Carlo simulations
of the OFDM-based RS-SCMA receiver, which employs SIC followed by MPA-based
private-layer detection. The system has $J=6$ users and $K=4$ REs per RS-SCMA
block. The common layer uses QPSK, and the private layer employs the SCMA
codebook in \cite{Li2022}, which is also used in \cite{RS_SCMA}. Each OFDM
symbol contains $N_{\mathrm{sc}}=128$ subcarriers, i.e., $Q=32$ RS-SCMA
blocks, and a CP of length $N_{\mathrm{cp}}=32$. The REs of each block are
mapped onto interleaved subcarriers. An eight-tap frequency-selective Rayleigh
fading channel with an exponential power-delay profile is considered
\cite{ofdm_scma}. Since the BER of user $j$ depends only on its own CFO
$\varepsilon_j$, the normalized CFO is set to be identical for all users,
$\varepsilon_j=\varepsilon\in\{0,0.02,0.05\}$, and the message-splitting factor
is varied as $\alpha\in\{0.25,0.5,0.75\}$. The main simulation parameters are
summarized in Table~\ref{tab:sim_params}.

\begin{table}[!htbp]
\centering
\footnotesize
\caption{Simulation Parameters}
\label{tab:sim_params}
\begin{tabular}{ll}
\toprule
Parameter & Value \\
\midrule
Number of users, $J$ & 6 \\
Number of REs per block, $K$ & 4 \\
Common-layer constellation & QPSK ($M=4$) \\
Private-layer SCMA codebook size, $M$ & 4 \\
Bits per block (common/private) & 8/12 \\
Number of subcarriers, $N_{\mathrm{sc}}$ & 128 \\
RS-SCMA blocks per OFDM symbol, $Q$ & 32 \\
CP length, $N_{\mathrm{cp}}$ & 32 \\
Subcarrier mapping & Interleaved \\
Number of channel taps, $L$ & 8 \\
Power-delay profile & Exponential \\
Normalized CFO, $\varepsilon$ & $\{0,\,0.02,\,0.05\}$ \\
Message-splitting factor, $\alpha$ & $\{0.25,\,0.5,\,0.75\}$ \\
Number of MPA iterations & 10 \\
\bottomrule
\end{tabular}
\end{table}

The max-min fairness (MMF) power allocation of \cite{RS_SCMA} progressively
reduces the private-layer power as the SNR increases. Under CFO, this
reduction, rather than the CFO itself, could then dominate the private-layer
BER degradation. To isolate the effect of CFO, the private-layer power
coefficient is held fixed once the estimated CFO-induced ICI power becomes
comparable to the noise power, and the remaining power is assigned to the
common layer so that the total transmit-power constraint is preserved. The
analytical results are evaluated using the expressions in
Section~\ref{sec:error_performance_analysis}, and the simulation results are
averaged over independent channel realizations under the same settings.
Performance is reported in terms of the common-layer BER
($\mathrm{BER}_{\mathrm c}$), the private-layer BER
($\mathrm{BER}_{\mathrm p}$), and the overall BER
($\mathrm{BER}_{\mathrm o}$).
    \def\datafile{pow_imb_mode1_v1.txt}   
\def\datafil{pow_imb_mode1_v1.txt}
\def\datafilcfo{pow_imb_mode1_v1.txt}
 \def\simfile{pow_imb_mode3_v1.txt}

\pgfplotsset{
  berstyle/.style={
    xmin=10, xmax=40,
    ymin=1e-5, ymax=6e-1,
    xtick={10,15,20,25,30,35,40},  
     tick label style={font=\footnotesize},
    xlabel={\footnotesize $E_b/N_0$ (dB)},
    ylabel={\footnotesize BER},
    grid=both, 
    grid style={dotted, gray!70},  
    legend cell align=left,width=1\linewidth,  
    height=7.5cm, 
    legend style={
      at={(0,0)},            
      anchor=south west,
      draw=none,               
      fill=none, 
      font=\scriptsize,
      legend columns=1,            
      column sep=1.5ex
    },
    every axis plot/.append style={thick, mark size=3pt}
  }
}

  
  


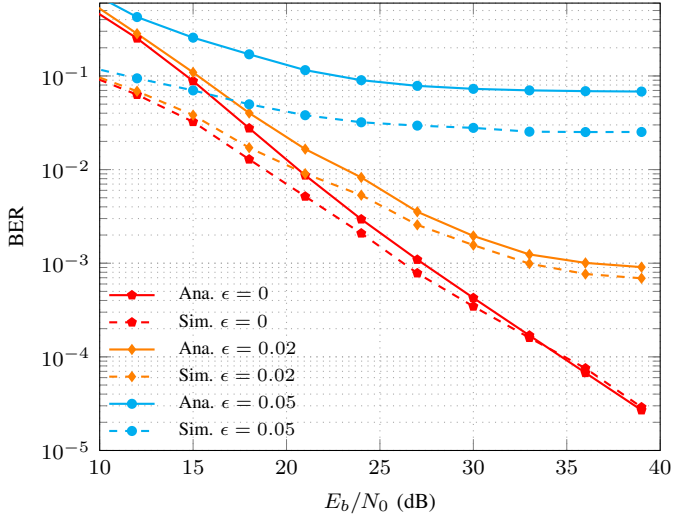
\begin{figure}[!htpb]
\centering
\begin{tikzpicture}
\begin{semilogyaxis}[berstyle]

  \addplot [red, mark=pentagon*, mark size=1.5pt]
  table [x=SNR, y=BERo_p00] {\datafil};
  \addlegendentry{Ana. $\epsilon=0$}

  \addplot [red, dashed, mark=pentagon*, mark size=1.5pt,
            mark options={solid}]
  table [x=SNR, y=BERo_p00] {\simfile};
  \addlegendentry{Sim. $\epsilon=0$}
  
  \addplot [orange, mark=diamond*, mark size=1.5pt]
  table [x=SNR, y=BERo_p02] {\datafilcfo};
  \addlegendentry{Ana. $\epsilon=0.02$}

  \addplot [orange, dashed, mark=diamond*, mark size=1.5pt,
            mark options={solid}]
  table [x=SNR, y=BERo_p02] {\simfile};
  \addlegendentry{Sim. $\epsilon=0.02$}
  
  \addplot [cyan, mark=oplus*, mark size=1.5pt]
  table [x=SNR, y=BERo_p05] {\datafile};
  \addlegendentry{Ana. $\epsilon=0.05$}

  \addplot [cyan, dashed, mark=oplus*, mark size=1.5pt,
            mark options={solid}]
  table [x=SNR, y=BERo_p05] {\simfile};
  \addlegendentry{Sim. $\epsilon=0.05$}

\end{semilogyaxis}
\end{tikzpicture}
\caption{\footnotesize Overall BER ($\mathrm{BER}_{\mathrm o}$) of the
OFDM-based RS-SCMA system ($J=6$, $K=4$) under CFO impairment for different
normalized CFO values $\varepsilon$.}
\label{fig:ber_overall_cfo}
\end{figure}

\pgfplotsset{
  berstyle/.style={
    xmin=10, xmax=40,
    ymin=8e-6, ymax=6e-1,
    xtick={10,15,20,25,30,35,40},
    tick label style={font=\footnotesize},
    xlabel={\footnotesize $E_b/N_0$ (dB)},
    ylabel={\footnotesize BER},
    grid=both,
    grid style={dotted, gray!70},
    legend cell align=left,
    every axis plot/.append style={
      thick,
      mark size=1.5pt
    },
    width=1.1\linewidth,
    height=8cm,
  }
}

\begin{figure*}[!htpb]
    \centering

    \begin{minipage}{0.42\textwidth}
        \centering
        \begin{tikzpicture}
        \begin{semilogyaxis}[
            berstyle,
            legend style={
              at={(1,1)},
              anchor=north east,
              draw=none,
              fill=none,
              font=\footnotesize,
              legend columns=1,
              column sep=1.5ex
            }
        ]

          \addplot [blue, mark=pentagon*]
          table [x=SNR, y=BERc_p00] {\datafil};
          \addlegendentry{Ana. $\epsilon=0$}

          \addplot [blue, dashed, mark=pentagon*,
                    mark options={solid}]
          table [x=SNR, y=BERc_p00] {\simfile};
          \addlegendentry{Sim. $\epsilon=0$}

          \addplot [magenta, mark=diamond*]
          table [x=SNR, y=BERc_p02] {\datafilcfo};
          \addlegendentry{Ana. $\epsilon=0.02$}

          \addplot [magenta, dashed, mark=diamond*,
                    mark options={solid}]
          table [x=SNR, y=BERc_p02] {\simfile};
          \addlegendentry{Sim. $\epsilon=0.02$}

          \addplot [green!60!black, mark=oplus*]
          table [x=SNR, y=BERc_p05] {\datafile};
          \addlegendentry{Ana. $\epsilon=0.05$}

          \addplot [green!60!black, dashed, mark=oplus*,
                    mark options={solid}]
          table [x=SNR, y=BERc_p05] {\simfile};
          \addlegendentry{Sim. $\epsilon=0.05$}

        \end{semilogyaxis}
        \end{tikzpicture}

        \vspace{1ex}
        \footnotesize (a) Common-layer $\mathrm{BER}_{\mathrm c}$
    \end{minipage}%
    \hspace{1.8cm}%
    \begin{minipage}{0.42\textwidth}
        \centering
        \begin{tikzpicture}
        \begin{semilogyaxis}[
            berstyle,
            legend style={
              at={(0,0)},
              anchor=south west,
              draw=none,
              fill=none,
              font=\footnotesize,
              legend columns=1,
              column sep=1.5ex
            }
        ]

          \addplot [violet, mark=pentagon*]
          table [x=SNR, y=BERp_p00] {\datafil};
          \addlegendentry{Ana. $\epsilon=0$}

          \addplot [violet, dashed, mark=pentagon*,
                    mark options={solid}]
          table [x=SNR, y=BERp_p00] {\simfile};
          \addlegendentry{Sim. $\epsilon=0$}

          \addplot [brown!45!black, mark=diamond*]
          table [x=SNR, y=BERp_p02] {\datafilcfo};
          \addlegendentry{Ana. $\epsilon=0.02$}

          \addplot [brown!45!black, dashed, mark=diamond*,
                    mark options={solid}]
          table [x=SNR, y=BERp_p02] {\simfile};
          \addlegendentry{Sim. $\epsilon=0.02$}

          \addplot [olive!85!black, mark=oplus*]
          table [x=SNR, y=BERp_p05] {\datafile};
          \addlegendentry{Ana. $\epsilon=0.05$}

          \addplot [olive!85!black, dashed, mark=oplus*,
                    mark options={solid}]
          table [x=SNR, y=BERp_p05] {\simfile};
          \addlegendentry{Sim. $\epsilon=0.05$}

        \end{semilogyaxis}
        \end{tikzpicture}

        \vspace{1ex}
        \scriptsize (b) Private-layer $\mathrm{BER}_{\mathrm p}$
    \end{minipage}

    \caption{\footnotesize Layer-wise BER of the OFDM-based RS-SCMA system
    ($J=6$, $K=4$) under CFO impairment: (a) common-layer
    $\mathrm{BER}_{\mathrm c}$ and (b) private-layer
    $\mathrm{BER}_{\mathrm p}$.}
    \label{fig:layerwise_ber}
\end{figure*}

Fig.~\ref{fig:ber_overall_cfo} compares the analytical and simulated overall
BER of the OFDM-based RS-SCMA system for
$\varepsilon\in\{0,0.02,0.05\}$. For $\varepsilon=0$, the analytical and
simulated results are in close agreement over the entire SNR range, and the
BER decreases continuously with $E_b/N_0$. For nonzero CFO, the BER gradually
approaches an error floor as $E_b/N_0$ increases. The onset and level of this
floor depend on the CFO magnitude: for $\varepsilon=0.02$, the curves begin
to saturate at relatively high $E_b/N_0$, whereas for $\varepsilon=0.05$,  the
saturation occurs earlier and at a considerably higher level of  BER.
For $\varepsilon=0.02$, the analytical result follows the simulation closely,
including the high-SNR floor of the order of $10^{-3}$. A larger
analytical--simulation gap is observed for $\varepsilon=0.05$, particularly
in the high-SNR region, where the analytical and simulated BERs approach
approximately $7\times10^{-2}$ and $2.5\times10^{-2}$, respectively.
Nevertheless, the analysis correctly captures the BER degradation with
increasing CFO and the resulting high-SNR saturation. The increasing
difference at larger CFO is mainly associated with the approximations used in
the BER evaluation, including the nearest-neighbour error events and the
Gaussian representation of the unresolved CFO-induced interference.

Fig.~\ref{fig:layerwise_ber} shows the common and private-layer BERs for
different CFO values. For $\varepsilon=0$, the analytical results closely
follow the simulations for both layers. With nonzero CFO, both layers develop
high-SNR error floors, with a much stronger degradation in the private layer.
For $\varepsilon=0.05$, the simulated private-layer BER remains on the order
of a few times $10^{-2}$, whereas the common-layer BER stays below
$10^{-3}$. The private  layer (SCMA layer)  carries $12$ bits per active block,
compared with $8$ bits for the common layer. Its higher BER and larger bit
weight therefore make it the dominant contributor to the overall BER.

Although both layers experience CFO-induced inter-RE coupling, their detection
conditions are different. The common layer is allocated higher transmit power
and is detected first in the presence of the private-layer signal. After
cancellation of the detected common symbols, the lower-power private SCMA
signal is detected using MPA. Thus, private-layer detection is affected by
CFO-induced interference and may also experience residual interference when
the common-layer decision is incorrect.
The analytical results closely follow the common-layer simulations, while the
analytical-simulation gap becomes more noticeable for the private layer as
the CFO increases. This gap mainly results from the nearest-neighbour
restriction in the BER evaluation and the Gaussian approximation of the
unresolved CFO-induced interference. The dominant-event approximation used for
SIC error propagation introduces an additional, but smaller, approximation.
Despite these approximations, the analysis captures the greater CFO
sensitivity of the private layer and its dominant contribution to the 
error floor.

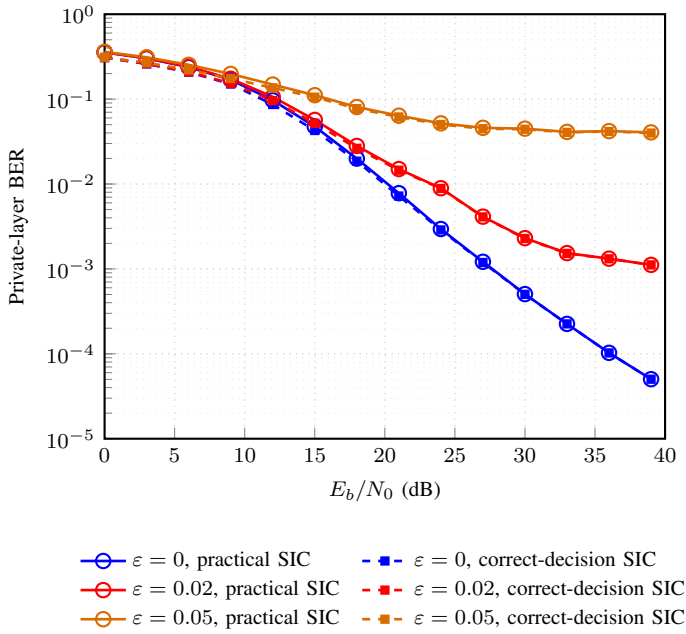
\begin{figure}[hbp!t]
\centering

\begin{tikzpicture}

\begin{semilogyaxis}[
    name=mainaxis,
    width=1\columnwidth,
    height=0.8\columnwidth,
    xmin=0,
    xmax=40,
    ymin=1e-5,
    ymax=1,
    xtick={0,5,10,15,20,25,30,35,40},
    ytick={1e-5,1e-4,1e-3,1e-2,1e-1,1},
    xlabel={$E_b/N_0$ (dB)},
    ylabel={Private-layer BER},
    tick align=inside,
    tick pos=left,
    tick label style={font=\footnotesize},
    label style={font=\footnotesize},
    axis line style={line width=0.8pt},
    grid=both,
    major grid style={dotted,gray!45},
    minor grid style={dotted,gray!18},
    legend style={
        at={(0.5,-0.24)},
        anchor=north,
        draw=none,
        fill=none,
        font=\footnotesize,
        legend columns=2,
        legend cell align=left,
        /tikz/every even column/.append style={column sep=6pt}
    }
]


\addplot[
    blue,
    solid,
    line width=1pt,
    mark=o,
    mark size=2.7pt,
    mark options={
        solid,
        fill=white,
        line width=0.8pt
    }
]
table[
    x=EbN0,
    y=eps00_imperfect,
    col sep=space
]{perfect_vs_imperfect_sic_data.txt};
\addlegendentry{$\varepsilon=0$, practical SIC}

\addplot[
    blue,
    dashed,
    line width=1pt,
    mark=square*,
    mark size=1.25pt,
    mark options={solid}
]
table[
    x=EbN0,
    y=eps00_perfect,
    col sep=space
]{perfect_vs_imperfect_sic_data.txt};
\addlegendentry{$\varepsilon=0$, correct-decision SIC}


\addplot[
    red,
    solid,
    line width=1pt,
    mark=o,
    mark size=2.7pt,
    mark options={
        solid,
        fill=white,
        line width=0.8pt
    }
]
table[
    x=EbN0,
    y=eps02_imperfect,
    col sep=space
]{perfect_vs_imperfect_sic_data.txt};
\addlegendentry{$\varepsilon=0.02$, practical SIC}

\addplot[
    red,
    dashed,
    line width=1pt,
    mark=square*,
    mark size=1.25pt,
    mark options={solid}
]
table[
    x=EbN0,
    y=eps02_perfect,
    col sep=space
]{perfect_vs_imperfect_sic_data.txt};
\addlegendentry{$\varepsilon=0.02$, correct-decision SIC}


\addplot[
    orange!85!black,
    solid,
    line width=1pt,
    mark=o,
    mark size=2.7pt,
    mark options={
        solid,
        fill=white,
        line width=0.8pt
    }
]
table[
    x=EbN0,
    y=eps05_imperfect,
    col sep=space
]{perfect_vs_imperfect_sic_data.txt};
\addlegendentry{$\varepsilon=0.05$, practical SIC}

\addplot[
    orange!85!black,
    dashed,
    line width=1pt,
    mark=square*,
    mark size=1.25pt,
    mark options={solid}
]
table[
    x=EbN0,
    y=eps05_perfect,
    col sep=space
]{perfect_vs_imperfect_sic_data.txt};
\addlegendentry{$\varepsilon=0.05$, correct-decision SIC}

\end{semilogyaxis}

\end{tikzpicture}

\caption{Private-layer BER with practical SIC and correct-decision SIC
for different normalized CFO values.}
\label{fig:perfect_vs_imperfect_sic}

\end{figure}
\begin{figure}[!t]
\centering
\begin{tikzpicture}
\begin{groupplot}[
    group style={group size=1 by 2, vertical sep=9pt},
    width=0.94\columnwidth,
    height=0.74\columnwidth,
    xmin=0, xmax=1,
    xtick={0,0.2,0.4,0.6,0.8,1},
    tick align=inside, tick pos=left, tick style={black},
    tick label style={font=\footnotesize},
    label style={font=\footnotesize},
    axis line style={line width=0.8pt}
]

\nextgroupplot[
    height=0.6\columnwidth,
    axis lines=box,
    ymin=100, ymax=260, ytick={100,150,200,250},
    xticklabels=\empty, xlabel=\empty,
    xticklabel style={font=\scriptsize},
    yticklabel style={font=\scriptsize},
    ylabel={$\lambda_{\mathrm{RS\text{-}SCMA}}$ (\%)},
    grid=major, grid style={dotted,gray!35},
    legend style={
        at={(0.5,0.06)}, anchor=south,
        draw=none, fill=white, fill opacity=0.7, text opacity=1,
        font=\scriptsize
    }
]
\addplot[green, solid, line width=1.1pt, no marks]
table[x=x, y=y, col sep=space]{lc_vs_lambda.txt};
\addlegendentry{Overloading factor}

\nextgroupplot[
    height=0.75\columnwidth,
    axis lines=box,
    ymode=log, log basis y=10,
    ymin=1e-6, ymax=1e-1,
    ytick={1e-6,1e-5,1e-4,1e-3,1e-2,1e-1},
    yticklabels={$10^{-6}$,$10^{-5}$,$10^{-4}$,$10^{-3}$,$10^{-2}$,$10^{-1}$},
    xticklabel style={font=\footnotesize},
    yticklabel style={font=\footnotesize},
    minor y tick num=9,
    grid=both,
    major grid style={dotted,gray!45},
    minor grid style={dotted,gray!18},
    xlabel={Message-splitting factor $\alpha$},
    ylabel={Overall BER},
    legend style={
        at={(0.5,0.04)}, anchor=south,
        draw=none, fill=white, fill opacity=0.7, text opacity=1,
        font=\scriptsize,
        legend columns=3, legend cell align=left,
        /tikz/every even column/.append style={column sep=6pt}
    }
]
\addplot[blue, solid, line width=1.0pt, mark=o, mark size=2.2pt,
    mark repeat=2, mark options={solid, fill=white}]
table[x=alpha, y=y00, col sep=space]{cfo_bervsalpha.txt};
\addlegendentry{$\varepsilon=0$}

\addplot[red, dashed, line width=1.0pt, mark=square*, mark size=2.2pt, mark repeat=2]
table[x=alpha, y=y02, col sep=space]{cfo_bervsalpha.txt};
\addlegendentry{$\varepsilon=0.02$}

\addplot[orange!85!black, dashdotted, line width=1.0pt, mark=triangle*,
    mark size=2.4pt, mark repeat=2]
table[x=alpha, y=y05, col sep=space]{cfo_bervsalpha.txt};
\addlegendentry{$\varepsilon=0.05$}

\end{groupplot}
\end{tikzpicture}
\caption{\footnotesize Effective RS-SCMA overloading factor (top) and overall BER (bottom)
versus the message-splitting factor $\alpha$ at
$E_b/N_0=30$~dB for different normalized CFO values.}
\label{fig:ber_vs_alpha}
\end{figure}
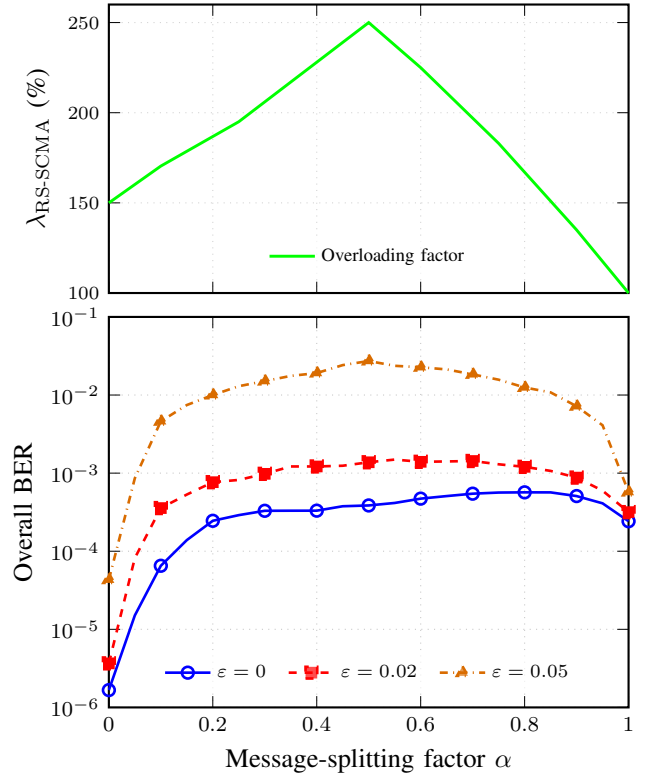

To examine the effect of common-decision errors,
Fig.~\ref{fig:perfect_vs_imperfect_sic} compares the private-layer BER under
practical SIC and correct-decision SIC. With practical SIC, cancellation is
performed using the detected common symbols, so erroneous common decisions
produce residual interference. With correct-decision SIC, the transmitted
common symbols are used for cancellation. This removes common-decision error
propagation while retaining the CFO-induced leakage.
The BER curves for the two SIC cases remain close over all considered CFO
values.  This shows that common-decision error propagation
has only a limited effect on the private-layer BER under the considered
settings. The persistent error floor is mainly associated with
CFO-impaired private-layer detection rather than SIC error propagation.
Fig.~\ref{fig:ber_vs_alpha} shows the effective RS-SCMA overloading factor~ \eqref{eq:rsscma_effective_overloading}
and the simulated overall BER as functions of the message-splitting factor
$\alpha$. The overloading factor depends only on the RS-SCMA transmission
structure and is independent of CFO. For the considered $J=6$ and $K=4$
configuration, it increases from $150\%$ at $\alpha=0$ to its maximum of
$250\%$ at $\alpha=0.5$, and then decreases to $100\%$ at $\alpha=1$.
Thus, equal message splitting provides the highest effective overloading,
since the common and private layers remain jointly active throughout the
transmission.

The lower panel of  Fig.~\ref{fig:ber_vs_alpha} shows the corresponding overall BER at
$E_b/N_0=30$~dB for $\varepsilon\in\{0,0.02,0.05\}$. For all considered
CFO values, the BER increases as $\alpha$ approaches $0.5$. At this point,
the transmission consists entirely of the joint common-private phase, so
both layers are simultaneously affected by CFO-induced interference and
private-layer detection may also be affected by common-layer decision errors
through SIC. Moving away from $\alpha=0.5$ introduces a private-only interval
for $\alpha<0.5$ or a common-only interval for $\alpha>0.5$, reducing the
fraction of jointly transmitted blocks.
The BER variation is asymmetric about $\alpha=0.5$. For $\alpha<0.5$, the
increasing private-only interval leads to a pronounced BER reduction as
$\alpha$ approaches zero. For $\alpha>0.5$, the BER decreases more gradually
over much of the common-dominant region and drops more noticeably only as
$\alpha$ approaches one. This difference reflects the distinct signaling and
detection structures of the private SCMA and common layers. The effect of CFO
is also evident from the increasing BER level as $\varepsilon$ grows,
particularly around the equal-splitting region.
The figure highlights the tradeoff between reliability and
effective overloading. Although the lowest BER is obtained near the
single-layer operating points, the maximum overloading occurs at
$\alpha=0.5$.

\def\bervseps{epsvsber.txt}

\pgfplotsset{
  berstyle/.style={
   width=1\columnwidth,
    height=0.75\columnwidth,
    xmin=0,
    xmax=0.10,
    ymin=1e-4,
    ymax=2e-1,
    xtick={0,0.02,0.04,0.06,0.08,0.10},
    xticklabels={0,0.02,0.04,0.06,0.08,0.10},
    tick label style={font=\footnotesize},
    label style={font=\footnotesize},
    xlabel={Normalized CFO, $\varepsilon$},
    ylabel={Overall BER},
    grid=both,
    grid style={dotted,gray!70},
    legend cell align=left,
    legend style={
        font=\footnotesize,
        at={(0.99,0.01)},
        anchor=south east,
        draw=none,
        fill=white,
        fill opacity=0.85,
        text opacity=1
    },
    every axis plot/.append style={
        thick,
        mark size=1.5pt
    },
    width=1\linewidth,
    height=7.5cm
  }
}
\begin{figure}[!t]
\centering
\begin{tikzpicture}
\begin{semilogyaxis}[
    berstyle
]

\addplot[
    blue,
    solid,
    mark=o,
    mark options={
        solid,
        fill=white
    }
]
table[
    x=eps,
    y=y01
]{\bervseps};
\addlegendentry{$\alpha=0.1$}

\addplot[
    cyan!70!black,
    dashed,
    mark=square,
    mark options={
        solid,
        fill=white
    }
]
table[
    x=eps,
    y=y03
]{\bervseps};
\addlegendentry{$\alpha=0.3$}

\addplot[
    red,
    solid,
    mark=square*,
    mark options={solid}
]
table[
    x=eps,
    y=y05
]{\bervseps};
\addlegendentry{$\alpha=0.5$}

\addplot[
    orange!90!black,
    dashdotted,
    mark=triangle*,
    mark options={solid}
]
table[
    x=eps,
    y=y07
]{\bervseps};
\addlegendentry{$\alpha=0.7$}

\addplot[
    magenta!80!black,
    densely dotted,
    mark=diamond*,
    mark options={solid}
]
table[
    x=eps,
    y=y09
]{\bervseps};
\addlegendentry{$\alpha=0.9$}

\end{semilogyaxis}
\end{tikzpicture}
\caption{Overall BER versus normalized CFO $\varepsilon$ for different
message-splitting factors at $E_b/N_0=30$ dB.}
\label{fig:ber_vs_cfo}
\end{figure}
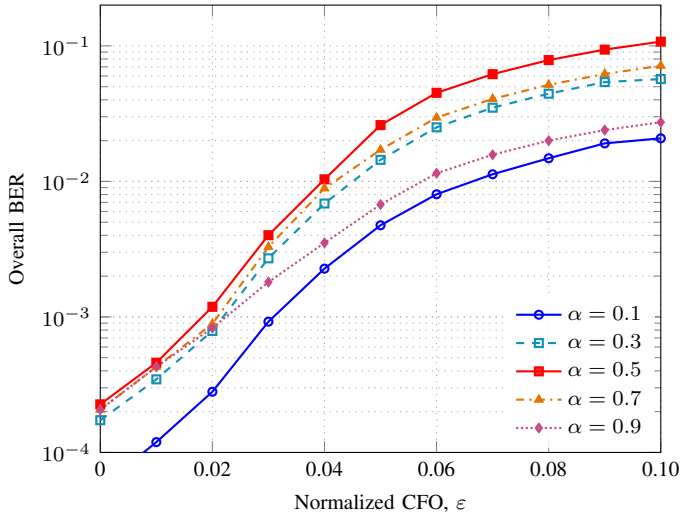
Fig.~\ref{fig:ber_vs_cfo} shows the simulated overall BER versus the
normalized CFO $\varepsilon$ at $E_b/N_0=30$~dB for different
message-splitting factors. For all values of $\alpha$, the BER increases with
$\varepsilon$ due to the increasing CFO-induced ICI. The degradation is mild
at small offsets and becomes more pronounced as the CFO increases.
The BER ordering depends on the relative importance of two effects. At low
CFO, the penalty associated with joint common--private transmission is small,
and the performance is mainly influenced by the single-layer detection
quality. Hence, the private-dominant cases, such as $\alpha=0.1$ and
$\alpha=0.3$, achieve lower BER, while $\alpha=0.9$ performs less favorably
because most of the transmission occurs in the common-only interval.
As the CFO increases, the joint transmission interval becomes increasingly
sensitive to ICI. The duration of this interval then becomes more important.
For example, $\alpha=0.9$ has a joint interval of only $0.1N$, compared with
$0.3N$ for $\alpha=0.3$ and $\alpha=0.7$. Consequently, the BER curve for
$\alpha=0.9$ crosses those of the latter cases at moderate CFO and becomes
lower at larger offsets. This crossover reflects the changing balance between
single-layer detection quality and the duration of CFO-affected joint
transmission.
\section{Conclusion}
\label{sec:conclusion}

This paper presented a finite-alphabet BER analysis of OFDM-based RS-SCMA systems operating in the presence of CFO over frequency-selective Rayleigh fading channels. The analysis accounted for the common and private signal layers,
CFO-induced interference, and SIC error propagation. Numerical results demonstrated that CFO significantly degrades BER performance and leads to a high-SNR error floor, with the floor level increasing with the normalized frequency offset. The private SCMA layer exhibited greater sensitivity to CFO and was identified as the dominant contributor to the overall BER degradation. The analytical results closely matched the simulation results over the considered CFO values and message-splitting factors. These results provide useful analytical insights into the impact of CFO on RS-SCMA systems and can serve as a basis for the design and performance evaluation of CFO-resilient RS-SCMA transceivers.


\balance
\bibliographystyle{IEEEtran}
\footnotesize
			\bibliography{references_ofdm}
\end{document}